\documentclass[journal]{IEEEtran}
\usepackage{amsmath,amsfonts}
\usepackage{algorithmic}
\usepackage{algorithm}
\usepackage{bm}
\usepackage{array}
\usepackage[caption=false,font=normalsize,labelfont=sf,textfont=sf]{subfig}
\usepackage{textcomp}
\usepackage{stfloats}
\usepackage{url}
\usepackage{verbatim}
\usepackage{graphicx}
\usepackage{cite}
\usepackage{subfig}
\usepackage{booktabs}
\usepackage{multirow}
\usepackage[T1]{fontenc}
\usepackage{mathtools}
\IEEEoverridecommandlockouts

\begin{document}

\title{ \LARGE \bf A Practical Large-Scale Roadside Multi-View Multi-Sensor Spatial Synchronization Framework for Intelligent Transportation Systems}

\author{Yong Li$^{1}$, Zhiguo Zhao$^{1}$, Yunli Chen$^{1}$, and Rui Tian$^{1}$$^{\dagger}$
\thanks{$^{1}$The authors are with the Faculty of Information Technology, Beijing University of Technology, Beijing, 100124, China (Rui Tian is the corresponding author. phone: +8618010027292; e-mail: {\tt\small rui.tian@bjut.edu.cn}).}
}

\maketitle
\thispagestyle{empty}
\pagestyle{empty}
\begin{abstract}
  Spatial synchronization in roadside scenarios is a prerequisite for the fusion of multi-source information obtained by various sensors at different locations.
  However, the current spatial synchronization methods based on the cascading spatial transformation (CST) can lead to considerable cumulative errors when deployed on a large scale. 
  Manual camera calibration, which is also commonly used, is insufficient to handle scenarios with changing camera parameters and requires substantial manual intervention when deployed extensively. 
  Additionally, existing methods are often limited to controlled or single-view scenarios, lacking consideration for various practical problems that arise in large-scale deployments. 
  To address these issues, this research introduces a customized parallel spatial transformation (PST)-based spatial synchronization framework designed for large-scale, multi-view, multi-sensor scenarios. 
  PST directly parallelizes the transformation of all sensor coordinate systems into global coordinate system  without accumulating errors in CST.
  Furthermore, For achieving precise roadside monocular global localization while considering scenarios with varying camera parameters and unknown camera heading angles, this research incorporates deep learning into camera calibration, distance and angle estimation, and automatic heading angle calculation, reducing manual intervention.
  Finally, in mitigating the potential reduction in synchronization accuracy in real-world settings, this paper utilizes geolocation cues and an optimization algorithm based on stochastic gradient descent to improve spatial synchronization precision.
  The proposed method has been systematically tested in real-world scenarios under different parameters and data conditions, and experimental results demonstrate its significant advantages over previous CST-based methods. 
  It can greatly improve the efficiency of large-scale roadside multi-perspective, multi-sensor spatial synchronization and reduce deployment costs.
\end{abstract}

\section{INTRODUCTION}
\IEEEPARstart{W}{ith} the increasing prevalence of intelligent transportation systems, the deployment of roadside units (RSUs) has notably surged, keeping pace with the growing adoption of intelligent transportation technologies\cite{Nguyen2022DynamicVC,Magsino2022AnEI}. 
Leveraging and integrating data from various multi-modal sensors, RSUs can provide a more comprehensive and extensive road traffic perception compared to onboard sensors, signifying a critical trajectory for the advancement of vehicle-to-everything (V2X) intelligent transportation services. 
In recent years, multi-sensor data fusion has gained significant traction in the fields of computer vision and intelligent transportation, involving the integration of data from diverse modalities such as millimeter-wave (MMW) radar, Light Detection and Ranging (LiDAR), camera, and ultrasonic sensor. 
According to research findings\cite{fuCameraRadarFusion2020, duNovelSpatioTemporalSynchronization2022}, ensuring accurate spatial synchronization of sensors is crucial for successful multi-sensor data fusion. 
This becomes particularly challenging for roadside multi-sensor fusion, given the diverse sensors locations. 
Furthermore, roadside sensors can be expanded, unlike the fixed setups of on-board sensors. 
However, during the expansion of sensors, conventional spatial synchronization methods may accumulate significant errors, emphasizing the necessity for comprehensive research on temporal and spatial synchronization for large-scale roadside sensors.
\begin{figure}[t]
  \centering
  \subfloat[]{\includegraphics[width=0.5\columnwidth]{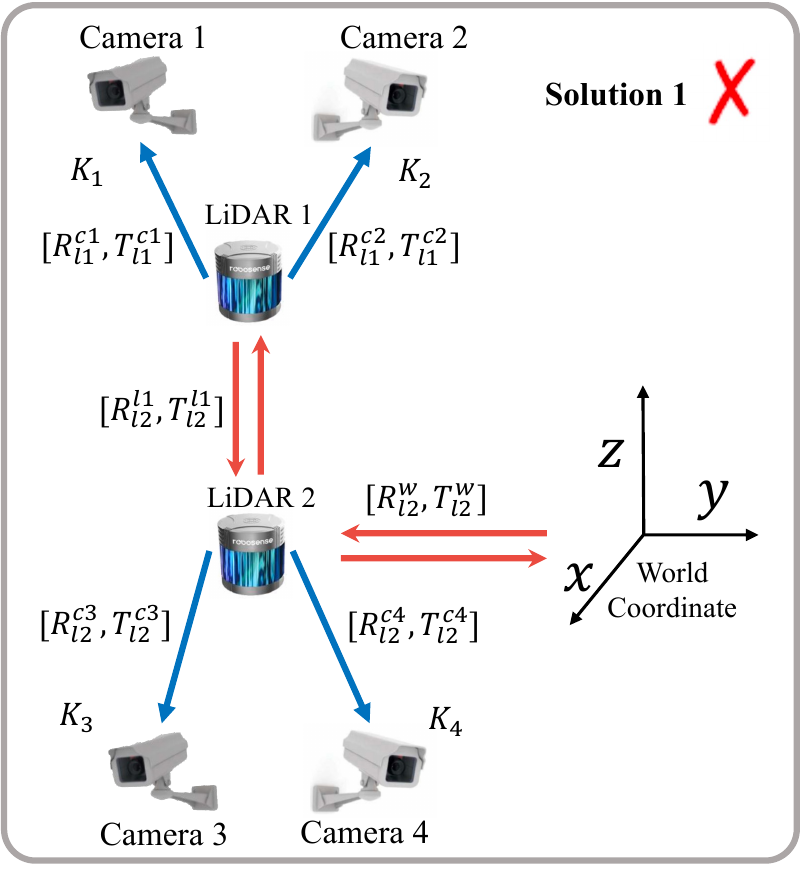}%
  \label{CST}}
  \hfil
  \subfloat[]{\includegraphics[width=0.5\columnwidth]{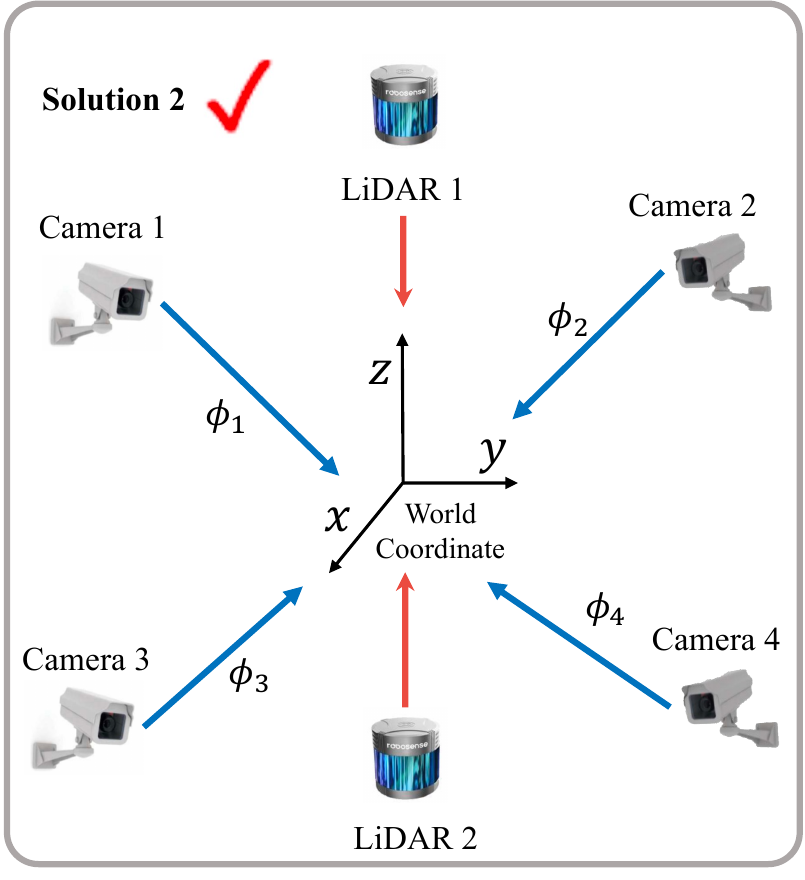}%
  \label{PST}}
  \caption{The comparison of previous method based on CST and our method based on PST. In the typical scenario of multi-sensor deployment at urban traffic intersections, our framework achieves spatial synchronization not through a cascading approach but by directly unifying all sensors or sensor groups into a global coordinate system. Additionally, we optimize the parameters involved in synchronization as a whole, eliminating the need for manual calibration. This design not only reduces cumulative errors but also minimizes human intervention, thereby enhancing the precision and efficiency of system deployment.}
  \label{compare}
  \end{figure}

Currently, there is a substantial research gap in the synchronization of multiple views and sensors in large-scale roadside deployments.
Most existing studies focuses on experimental setups, as exemplified by Zheng $\textit{et al}$.'s work \cite{Zheng2022ARS}, which primarily explores controlled scenarios \cite{wangTimeSynchronizationSpace2023}, or single-view scenarios as discussed by Du $\textit{et al}$. \cite{duNovelSpatioTemporalSynchronization2022}.
Roadside spatial synchronization typically involves the CST, as shown in Fig. \ref{compare}\ref{sub@CST}, which includes steps such as pairing sensors to form groups within a relative coordinate system. 
This process gradually integrates new sensors with previous ones, culminating in the calibration of the last sensor group to the global world coordinate system, establishing a unified coordinate system.
Notably, before joint calibration, the camera and radar must be calibrated separately in advance.
These methods \cite{Zheng2022ARS, wangTimeSynchronizationSpace2023,duNovelSpatioTemporalSynchronization2022} assume that sensor calibration has been manually conducted, yielding either accurate or preliminary results.

While suitable for simpler experimental scenarios or on-board in autonomous driving, these methods all share the limitation of depending on operations within the existing coordinate system, leading to potential error accumulation and reduced spatial synchronization accuracy.
Manual intervention in large-scale roadside multi-view multi-sensor spatial synchronization scenarios results in inefficiencies and limited adaptability when simultaneously calibrating multiple cameras and radars (or LiDARs). 
Additionally, traditional manual camera calibration methods \cite{zhangFlexibleNewTechnique2000} have become inadequate with the increasing prevalence of Pan-Tilt-Zoom (PTZ) cameras and the potential for camera parameter changes due to external environmental factors.
Current automatic roadside camera calibration techniques, rooted in vanishing points, often rely on handcrafted features that are susceptible to false edges and excessively hinge on the precision of vanishing points.
While deep learning-based methods \cite{HoldGeoffroy2017APM,Zhu2020SingleVM,kocabasSPECSeeingPeople2021} can directly infer camera parameters from scenes, they may suffer from accuracy reduction due to domain shift in real-world scenarios.
Furthermore, both geometric and semantic-based calibration methods overlook an inherent feedback loop that ensures accurate calibration for specific downstream tasks.

To address the challenges of large-scale deployment, we introduce a novel approach termed PST, enabling concurrent spatial synchronization in a global coordinate system. 
By combining camera self-calibration, distance and angle estimation between the vehicle and the camera, and automatic estimation of the camera's heading angle, the approach achieves monocular global vehicle localization even in scenarios with unknown camera heading angles. 
To improve accuracy in real-world scenarios, we integrate LiDAR and camera GNSS (Global Navigation Satellite System) data to optimize localization and enhance resistance to interference. 
Additionally, utilizing GNSS coordinates in the calibration establishes a feedback loop, ensuring reliable localization.
To the best of our knowledge, we are the first to introduce the challenge of roadside multi-view multi-sensor spatial synchronization and have implemented a comprehensive and practical framework that is specifically designed to tackle this intricate problem.

This research produced the following major contributions:
\begin{enumerate}
  \item{A practical framework for large-scale roadside multi-sensor spatial synchronization, unifying sensor coordinates in a global system and achieving parallel spatial synchronization to minimize manual intervention and cumulative errors.}
  \item{Development of an automatic algorithm for roadside monocular localization, integrating deep learning for camera calibration, distance and angle estimation, and automatic camera heading angle estimation.}
  \item{Implementation of an optimization algorithm based on gradient descent to enhance monocular localization accuracy and mitigate spatial synchronization errors arising from global camera parameters and camera calibration inaccuracies caused by domain shift in real scenes.}
  \item{Conducting comprehensive real-world experiments to evaluate the framework's accuracy under varying parameter settings and data requirements.}
  \end{enumerate}

The rest of this paper is organized as follows. 
Section \ref{related work} provides a literature review,
Section \ref{problem statement and framework} introduces the problem statement and framework, 
Section \ref{monocular localization} presents the monocular localization algorithm, 
and Section \ref{spatial synchronization} outlines the spatial synchronization algorithm. 
The experimental results are discussed in Section \ref{experimental results}, while Section \ref{discussion} provides an analysis of the system's benefits, drawbacks, and future work. 
Finally, Section \ref{conclusion} concludes the paper.

\section{RELATED LITERATURE}\label{related work}
\subsection{Camera Calibration}
Camera calibration is crucial for tasks such as 3D reconstruction, precise measurements, and speed estimation in traffic scenes.
Conventional approaches like Direct Linear Transformation (DLT) \cite{AbdelAziz2015DirectLT} and Zhang's calibration method \cite{zhangFlexibleNewTechnique2000} often entail manual interventions, posing challenges in terms of scalability.
Recently emerging deep learning-based approaches \cite{Workman2015DEEPFOCALAM, Kendall2015PoseNetAC, HoldGeoffroy2017APM, Zhu2020SingleVM, kocabasSPECSeeingPeople2021} rely on extensive datasets but may struggle in complex scenes, leading to reduced accuracy.
Vanishing point-based automatic calibration is a favored method for roadside cameras \cite{Dubsk2015FullyAR}, \cite{Sochor2017TrafficSC} due to its adaptability to different camera types and mounting heights, but it depends on handcrafted features and precise vanishing point estimation, limiting its robustness.

Considering the complexity of these challenges, we adopts a deep learning-based method for roadside camera calibration. 
While it may sacrifice some precision in real-world scenarios, our approach meet rough localization requirements without scene context constraints and eliminates the need for manual curation of edge background models, offering an end-to-end camera parameter estimation solution.
\subsection{Roadside Monocular Vehicle Localization}
Monocular vehicle localization algorithms can be broadly categorized into two groups: those based on homography matrix mapping \cite{AbdelAziz2015DirectLT}, \cite{luCAROMVehicleLocalization2021} and those based on distance estimation \cite{rezaeiRobustVehicleDetection2015, Liu2017AND, Namazi2022GeolocationEO}.
The former typically necessitates the knowledge of specific coordinates to establish a mapping from image coordinates to world coordinates, and also often relying on intricate parameter adjustments, thus complicating large-scale deployments.
The latter method is commonly employed for on-board localization and relies on straightforward distance estimation principles suitable for practical scenarios.
Distance estimation methods commonly used include those based on IPM (Inverse Perspective Mapping) \cite{Wongsaree2018DistanceDT, Tuohy2010DistanceDF, Kim2019EfficientVD}, prior bounding box information \cite{Namazi2019UsingVC}, and camera geometric models \cite{rezaeiRobustVehicleDetection2015}.
Occasionally, these methods are combined to compensate for their respective limitations \cite{Liu2017AND, Namazi2022GeolocationEO}.
These distance estimation algorithms often depend on known camera parameters, which is not feasible for roadside-installed cameras. 
Such cameras may possess measurement errors upon installation or could be PTZ (Pan-Tilt-Zoom) cameras themselves, leading to parameter changes during operation.
However, previous methods have frequently relied on pre-existing parameters, selectively calibrated specific parameters, or entirely disregarded this crucial aspect, often overlooking the potential variations in camera parameters.
Moreover, distance estimation algorithms relying on IPM may introduce additional conversion errors, specifically through pixel mapping and interpolation, resulting in estimation errors of up to $\varepsilon=\pm8$, further compounded by inaccuracies during the calibration phase \cite{rezaeiRobustVehicleDetection2015}. 
For distance estimation based on prior bounding box information \cite{Namazi2019UsingVC}, accuracy heavily depends on the precision of prior vehicle size information and the detector, rendering them less effective when confronted with significant variation in vehicle size distribution.
Ojala $\textit{et al}$. \cite{Ojala2020NovelCN} proposed a pedestrian localization method, but their approach was conducted under indoor laboratory conditions, where cameras were calibrated using a chessboard pattern. 
However, in large-scale traffic scenarios, employing standard calibration methods such as chessboards for each sensor becomes cost-prohibitive and can even impede traffic, further complicating its practicality.
For the camera geometric model-based distance estimation methods, the existing work primarily focuses on longitudinal distance, with limited consideration for lateral distance, potentially leading to a certain degree of localization error \cite{rezaeiRobustVehicleDetection2015, Namazi2022GeolocationEO}.

Compared to existing localization algorithms, our monocular vehicle localization algorithm is tailored for large-scale roadside deployment, removing the necessity for known corresponding point coordinates, manual interventions, and prior information of vehicle size. 
Furthermore, it accommodates both lateral and longitudinal distances without relying on object detection model accuracy or scene constraints.
\subsection{Multi-Sensor Spatial Synchronization}
Existing spatial synchronization methods primarily rely on sequential one-to-one calibration, requiring extensive manual intervention \cite{duNovelSpatioTemporalSynchronization2022, fuCameraRadarFusion2020, wangRoadsideCameraRadarSensing2021}.
For instance, In the method proposed in \cite{Zheng2022ARS}, a set of sensors, including one LiDAR and two cameras, undergoes internal calibration followed by an external calibration with another set of sensors, leading to significant accumulated errors.
Similarly, approaches in \cite{fuCameraRadarFusion2020, wangTimeSynchronizationSpace2023} employ Zhang's method \cite{zhangFlexibleNewTechnique2000} for camera calibration.
Du $\textit{et al}$. \cite{duNovelSpatioTemporalSynchronization2022}, initially compute the homography matrix based on four corresponding points for camera calibration, followed by a spatio-temporal synchronization optimization model.
While these methods demonstrate high accuracy in individual settings, they encounter challenges in large-scale deployment scenarios.
In \cite{wangRoadsideCameraRadarSensing2021}, camera calibration is not performed directly. 
Instead, they combine a vast amount of correlated MMW radar and video object pairs and employ least-squares fitting or interpolation methods to estimate the spatial synchronization relationship function between the two coordinate systems.
While \cite{wangRoadsideCameraRadarSensing2021} achieves automation, the function fitting relies on substantial point-to-point data and is susceptible to various factors, including data quality, choice of fitting functions, and data distribution.

In contrast, our approach adopts a unified global coordinate system, enabling parallel spatial synchronization with minimal manual intervention. 
Additionally, Leveraging geolocation as a cue for spatial synchronization, our method minimizes the need for extensive manual intervention during the integration of new devices.
\section{PROBLEM STATEMENT AND PROPOSED FRAMEWORK}\label{problem statement and framework}
\begin{figure*}[t]
  \centering
  \includegraphics[width=\textwidth]{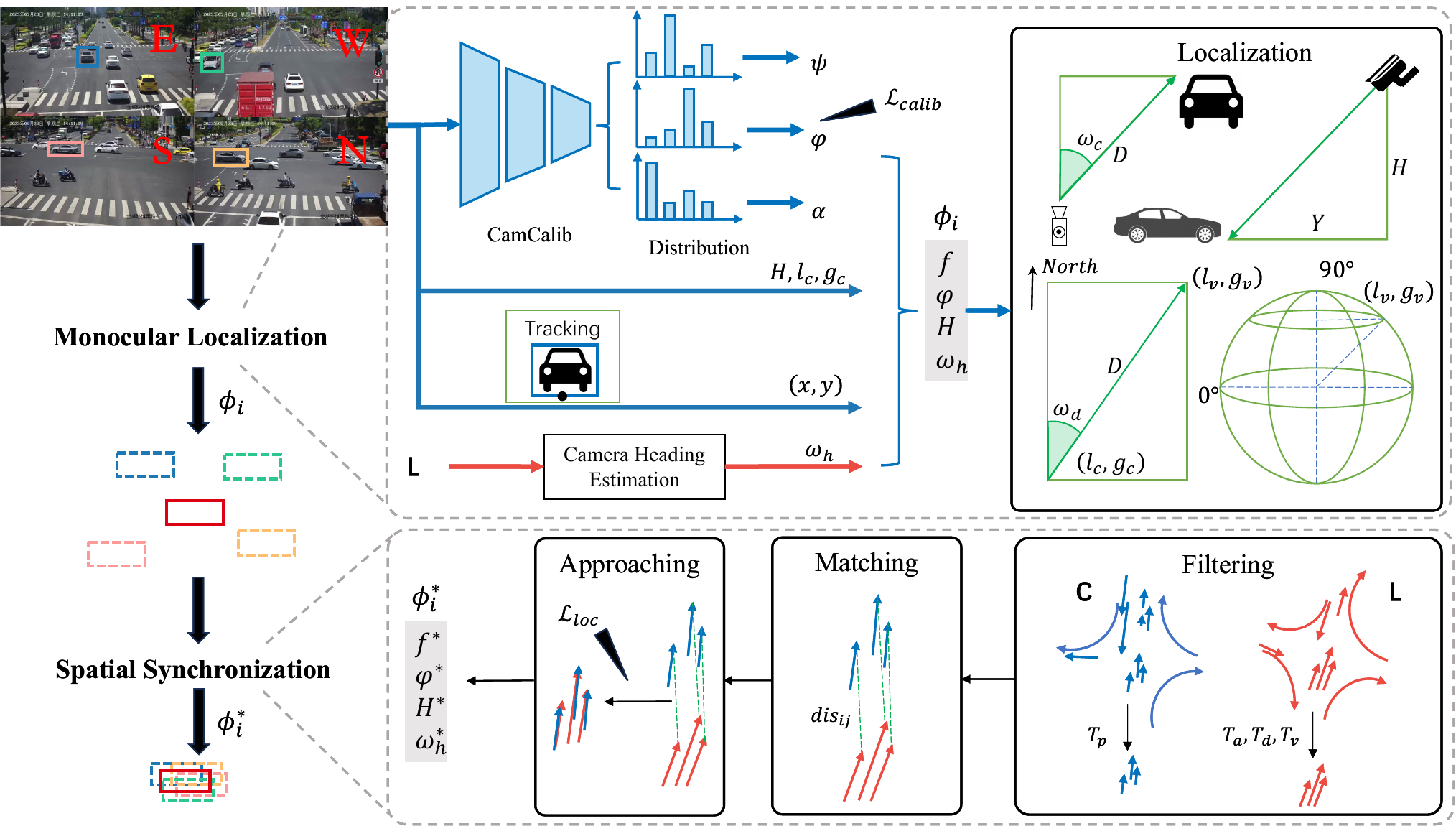}
  \caption{Overview of our framework. To achieve spatial synchronization in a global coordinate system, the system comprises two main modules: monocular localization and spatial synchronization. Given input images and LiDAR data from multiple perspectives, let's consider a single camera. Camera calibration, estimation of camera heading angles, and vehicle tracking are prerequisites for achieving global localization. After projecting pixel coordinates into the global GNSS, data from LiDAR and monocular global localization results are filtered. This process enables high-precision matching, and ultimately, optimization is performed on the matched data to regress the parameters. These parameters are then used to achieve precise global synchronization.}
  \label{framework_all}
  \end{figure*}
our goal is to achieve large-scale roadside multi-view, multi-sensor spatial synchronization by PST efficiently, which achieved by directly integrating all sensors into a global coordinate system, utilizing the GNSS coordinate system as the natural global reference. 
PST simplifies sensor integration, reducing time complexity and minimizing cumulative errors by enabling new sensors to synchronize with the global coordinate system without the need for complex considerations related to previously installed sensors.

To establish a unified global spatial coordinate system for various sensors, it is imperative that each sensor possesses GNSS localization capability.
LiDAR sensors seamlessly incorporate this functionality, and numerous manufacturers have already implemented GNSS localization when deploying these sensors at the roadside.
The primary challenge in achieving global localization for roadside monocular cameras depends on accurately mapping pixel coordinates to the spatial coordinate system, a process depends on exact camera calibration and the camera's global parameters (e.g., mounting height, heading angles, and geographic coordinates).
However, precise camera calibration and accurate global parameters is often unattainable, particularly when camera parameters change dynamically or measurement errors in position and angles can occur due to human error in large-scale deployment scenarios.
Therefore, optimizing these parameters becomes a necessity to achieve high-precision spatial synchronization, with the exception that the camera's latitude and longitude remain fixed values throughout the optimization process.

Let's assume the parameter combination for the global geolocation of the $i$-th camera is denoted as $\phi_i$. 
For a given scenario with $M$ vehicles, the world coordinates (i.e., obtained from the LiDAR geolocation) of the $j$-th vehicle are represented as $P_j$, and its pixel coordinates in the i-th camera are denoted as $p_{i_j}$. 
The objective is to find the optimal parameter combination $\phi_i^*$ that, when applied to the localization function $L$, minimizes the Euclidean distance error between the world coordinates $L_{\phi_i}(p_{i_j})$ projected from the pixel coordinates $p_{i_j}$ and the actual world coordinates $P_j$,
as expressed:
\begin{equation}
  \label{eq:problem 1}
  \phi_i^* = \text{argmin}_{\phi_i} \sum_j^{M} ||P_j - L_{\phi_i}(p_{i_j})||_2
\end{equation}

In a scenario with a total of $N$ cameras, the primary goal is to determine the optimal parameter combination $\Phi^*$ for these cameras. 
This optimal combination should minimize the cumulative Euclidean distance errors between the world coordinates $L_{\phi_i}(p_{i_j})$ derived by each camera using its specific parameter set $\phi_{i}$ and the actual vehicle coordinates $P_j$. 
This can be formulated as the following optimization problem:
\begin{equation}
  \label{eq:problem 2}
  \Phi = \{ \phi_1,\phi_1 \dots \phi_N \},
\end{equation}
\begin{equation}
  \label{eq:problem 3}
  \Phi^* = \text{argmin}_{\Phi} \sum_i^N \sum_j^M ||P_j - L_{\phi_{i}}(p_{i_j})||_2.
\end{equation}

PST proposed in this paper to address this problem is to independently solve for the optimal $\phi_i^*$ for each camera. 
Consequently, the final optimal parameter combination $\Phi^*$ is a collection of these individual solutions: $\Phi^* = \{ \phi_1^*,\phi_2^* \dots \phi_N^* \}$.

The framework proposed in this paper is given in Fig. \ref{framework_all}. 
Taking a typical traffic scenario as an example, cameras are typically installed in the east, west, south, and north directions. 
This installation includes other LiDAR or radar sensors. 
Within the framework, the addition of more LiDAR or cameras is treated as distinct groups for ease of analysis. 
Each video stream individually performs camera calibration, monocular localization, vehicle detection, and tracking. 
Camera heading angle estimation is unified within the LiDAR group, combining the camera's own latitude and longitude information for inference. 
Spatial synchronization is achieved by combining the vehicle's LiDAR tracking trajectories with the monocular GNSS trajectories of various cameras.
\section{MONOCULAR LOCALIZATION}\label{monocular localization}
\subsection{Camera Calibration}\label{sec:Camera Calibration}
\subsubsection{Projection Model}
\begin{figure}[t]
  \centering
  \subfloat[]{\includegraphics[width=0.55\columnwidth]{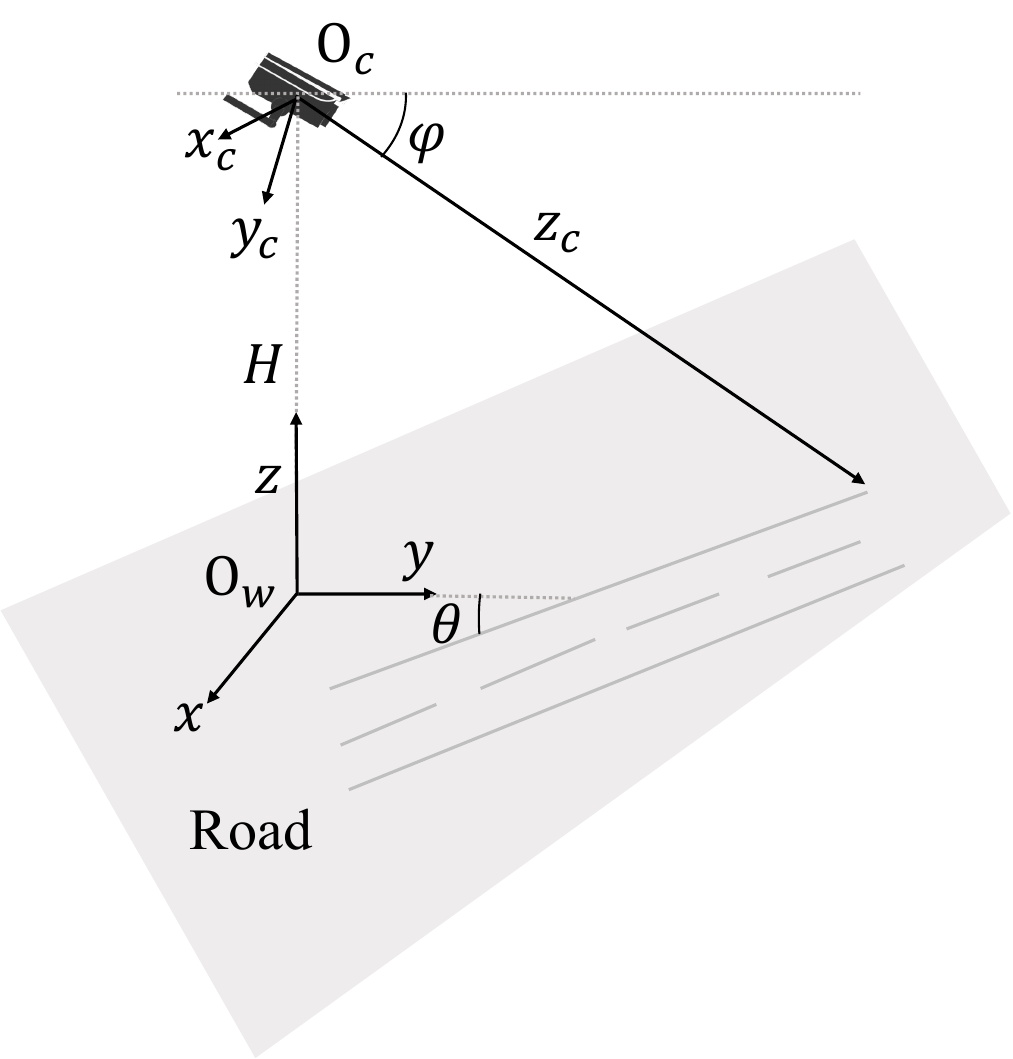}%
  \label{left view}}
  \hfil
  \subfloat[]{\includegraphics[width=0.4\columnwidth]{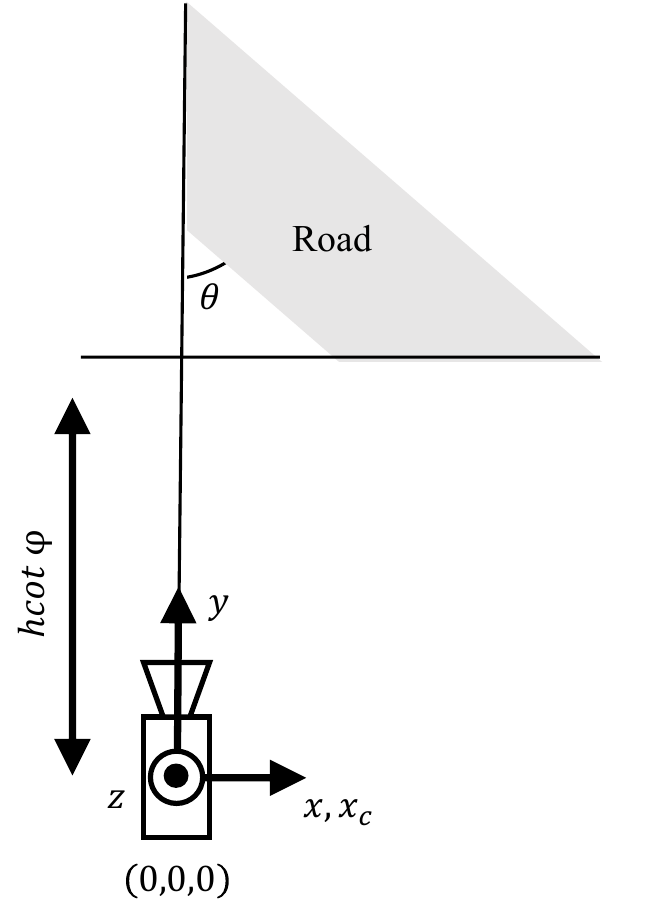}%
  \label{right view}}
  \caption{The mathematical relations between $\omega_h$, $\omega_d$, and $\omega_c$, along with a comparison of (a) vehicle position on the right side and (b) vehicle position on the left side within the camera's field of view.}
  \label{Camera model}
  \end{figure}
Consider a pinhole camera viewing a straight, flat road, a principal point at the image center, the camera has no skew and equal scaling in both horizontal and vertical directions as in [14], [15],\cite{Kanhere2010ATA}.
With these assumptions, exactly three parameters are needed to map Euclidean points on the road plane to image points on the image plane: the camera focal length $f$ , the height $H$ of the center of projection above the road plane, and the pitch angle $\phi$.
In addition, the yaw angle $\theta$ is needed to align the road coordinates with the direction of traffic flow. 
Since the roll angle of the camera in most roadside scenarios are typically very small, the errors in downstream tasks such as localization and speed estimation is negligible.

Fig. \ref{Camera model} illustrates the schematic diagram of the camera spatial model for road scenarios, with a right-handed coordinate system. 
As shown in Fig. \ref{Camera model}\ref{sub@left view}, this model defines the CCS $x_cy_cz_c$ and the WCS $xyz$. The CCS's origin is located at the principal point on the image plane, with the $x_c$ and $y_c$ axes aligned with the image rows and columns, respectively, and the $z_c$ axis aligned with the camera's optical axis.
The WCS has its origin aligned with the projection of the CCS's origin onto the ground plane, with the $x$ axis parallel to $x_c$ and the $y$ axis perpendicular to $x$ and lying on the road plane.
Considering the homogeneous representation of a point in 3D space as $\bm{x} = \begin{bmatrix} x, y, z, 1 \end{bmatrix}^T$, its transformation in the image coordinate system is $\bm{p} = \begin{bmatrix} s u, s v, s \end{bmatrix}^T$.
Here, $s \neq 0$, and thus, the mapping from the WCS $\bm{x}$ to the ICS $\bm{p}$ is given by:
\begin{equation}
  \label{eq:p}
  \bm{p} = \bm{K}\bm{R}\bm{T}\bm{x},
  \end{equation}
where $\bm{K}$, $\bm{R}$, and $\bm{T}$ respectively represent the intrinsic matrix, rotation matrix, and translation matrix. From the above analysis, it can be deduced that:
  \begin{equation}
    \label{eq:K}
    \bm{K} = 
    \begin{bmatrix}
    f & 0 & 0 \\
    0 & f & 0 \\
    0 & 0 & 1 \\
    \end{bmatrix},
    \end{equation}
    \begin{equation}
      \label{eq:R}
      \bm{R} = 
      \begin{bmatrix}
      1 & 0 & 0 \\
      0 & -\sin{\varphi} & -\cos{\varphi} \\
      0 & \cos{\varphi} & -\sin{\varphi} \\
      \end{bmatrix},
      \end{equation}
      \begin{equation}
        \label{eq:T}
        \bm{T} = 
        \begin{bmatrix}
        1 & 0 & 0 & 0\\
        0 & 1 & 0 & 0\\
        0 & 1 & 1 & -H\\
        \end{bmatrix}.
        \end{equation}  
Substituting equations \eqref{eq:K},\eqref{eq:R},\eqref{eq:T} into equation \eqref{eq:p} yields the expanded projection model.
\begin{equation}
  \label{eq:krt}
  \begin{bmatrix}
  s u \\
  s v \\
  s 
  \end{bmatrix}
   = 
  \begin{bmatrix}
  f & 0 & 0 & 0\\
  0 & -f\sin{\varphi} & -f\cos{\varphi} & fH\cos{\varphi}\\
  0 & \cos{\varphi} & -\sin{\varphi} & H\sin{\varphi}\\
  \end{bmatrix}
  \begin{bmatrix}
  x \\
  y \\
  z \\
  1 
  \end{bmatrix}.
  \end{equation}
\subsubsection{Network Architecture}
According to \cite{HoldGeoffroy2017APM, kocabasSPECSeeingPeople2021}, we discretize the  pitch $\varphi$, roll $\psi$, and vfov $\alpha$ angles into 256 bins in the spatial domain, so as to convert the regression problem into a B-way classification problem. 
We utilize ResNet-50 \cite{He2015DeepRL} as the backbone network and employ separate fully connected layers to predict the pitch angle $\varphi$, roll angle $\psi$, and vfov $\alpha$. 
The center values of each of the bins can be denoted as $\boldsymbol{\varphi} = [\varphi_1, . . . \varphi_i, . . . \varphi_B]$, $\boldsymbol{\psi} = [\psi_1, . . . \psi_i, . . . \psi_B]$, and $\boldsymbol{\alpha} = [\alpha_1, . . . \alpha_i, . . . \alpha_B]$, and let $\boldsymbol{p}^\varphi = [p_1^\varphi, . . . p_i^\varphi, . . . p_B^\varphi]$, $\boldsymbol{p}^\psi = [p_1^\psi, . . . p_i^\psi, . . . p_B^\psi]$, and $\boldsymbol{p}^\alpha = [p_1^\alpha, . . . p_i^\alpha, . . . p_B^\alpha]$ represent the probability mass from the fully-connected layers of each head respectively. 
The expected value of the probability mass is calculated as follows: 
\begin{equation}
  \begin{aligned}
  & \hat{\varphi}=\sum_i  p_i^\varphi\varphi_i, \\
  & \hat{\psi}=\sum_i p_i^\psi \psi_i, and \\
  & \hat{\alpha}=\sum_i p_i^\alpha \alpha_i,
  \end{aligned}
  \end{equation}
the difference between the predicted values and the ground truth is measured using the $\mathcal{L}_2$ loss. 
These operations effectively address the challenges associated with regression in a continuous space.

Since the range of the focal length (in pixels) is infinite and it changes when we adjust the image size, we estimate the vertical field of view (vfov) $\alpha$ in radians and convert it to focal length using the following equation, where $h$ represents the image pixel height:
\begin{equation}
  \label{eq:h, f}
  f=\frac{h}{2 \cdot \tan \left(\frac{\alpha}{2}\right)}.
\end{equation}
Since there is no natural frame of reference to estimate yaw (left vs right) from arbitrary images, we do not estimate yaw in our case. 
We can use the horizon line as an intuitive representation for these angles, $\varphi$ is indicated by the intersection of the horizontal midpoint and the vertical center line of the image, $\psi$ is determined by the rotation angle of the reference line in relation to the horizontal line.
\subsubsection{Loss Function}
As pointed out by \cite{kocabasSPECSeeingPeople2021}, predicting larger focal lengths compared to the ground truth (or equivalent to smaller vfov) yields better calibration results than predicting smaller focal lengths (larger vfov). Therefore, for pitch $\varphi$ and roll $\psi$ angles, we utilize the standard $\mathcal{L}_2$ loss between the prediction and the ground truth. For the vfov $\alpha$, we incorporate the Geman-McClure function \cite{Geman1987StatisticalMF}, applying a reduced penalty when the predicted vfov $\hat{\alpha}$ is smaller than the ground truth $\alpha$.
\begin{equation}
\mathcal{L}(\hat{\alpha})= \begin{cases}\frac{(\hat{\alpha}-\alpha)^2}{(\hat{\alpha}-\alpha)^2+1}, & \text { if } \hat{\alpha}<=\alpha \\ (\hat{\alpha}-\alpha)^2, & \text { otherwise.}\end{cases}
\end{equation}
The total loss is defined as follows:
\begin{equation}
\mathcal{L}_{calib}(\hat{\varphi}, \hat{\psi}, \hat{\alpha})=(\hat{\varphi}-\varphi)^2+(\hat{\psi}-\psi)^2+\mathcal{L}(\hat{\alpha}).
\end{equation}
\subsection{Camera Heading Estimation}
\begin{algorithm}[!b]
  \caption{Automatic Heading Angle Estimation}
  \label{algo:vehicle-selection-heading}
  \begin{minipage}[t]{0.8\textwidth}
    \textbf{Input:} $T_{h_d}$, $T_{t}$, $a_i $ and $ b_i$ for $i \in \{1, 2, \ldots, n\}$ \\
    \textbf{Output:} Most frequent heading angle $\Theta_{\text{mode}}$
  \end{minipage}
\begin{algorithmic}[1]
  \STATE \textbf{Initialize:} retained vehicle IDs list $R$
  \STATE \textbf{Initialize:} heading angles list $H$
  \FOR{$i$ from 1 to $n$}
  \IF{$CheckStraight$($a_i, b_i$)}
      \STATE $T_{\text{static}} \leftarrow StaticFind(i) \hfill \triangleright$ Stationary detection
      \IF{$T_{\text{static}} > T_{t}$} 
          \STATE $d \leftarrow DistanceCalculate(i, T_{\text{static}})$ 
          \IF{$d \leq T_{h_d}$}
              \STATE $R \leftarrow R \cup \{i\}$ \hfill $\triangleright$ Retain vehicle ID
              \STATE $\theta_h \leftarrow HeadingCalculate(a_i,b_i)$ \hfill $\triangleright$ Heading calculate 
              \STATE $H \leftarrow H \cup \{\theta_h\}$ \hfill $\triangleright$ Store $\theta_h$ in list $H$
          \ENDIF
      \ENDIF
  \ENDIF
\ENDFOR
\STATE $\Theta_{\text{mode}} \leftarrow \text{mode}(H)$ \hfill $\triangleright$ Calculate the mode (most frequent) heading angle
\RETURN Most frequent heading angle $\Theta_{\text{mode}}$
  \end{algorithmic}
\end{algorithm}
\begin{figure}[!t]
  \centering
  \subfloat[]{\includegraphics[width=0.34\columnwidth]{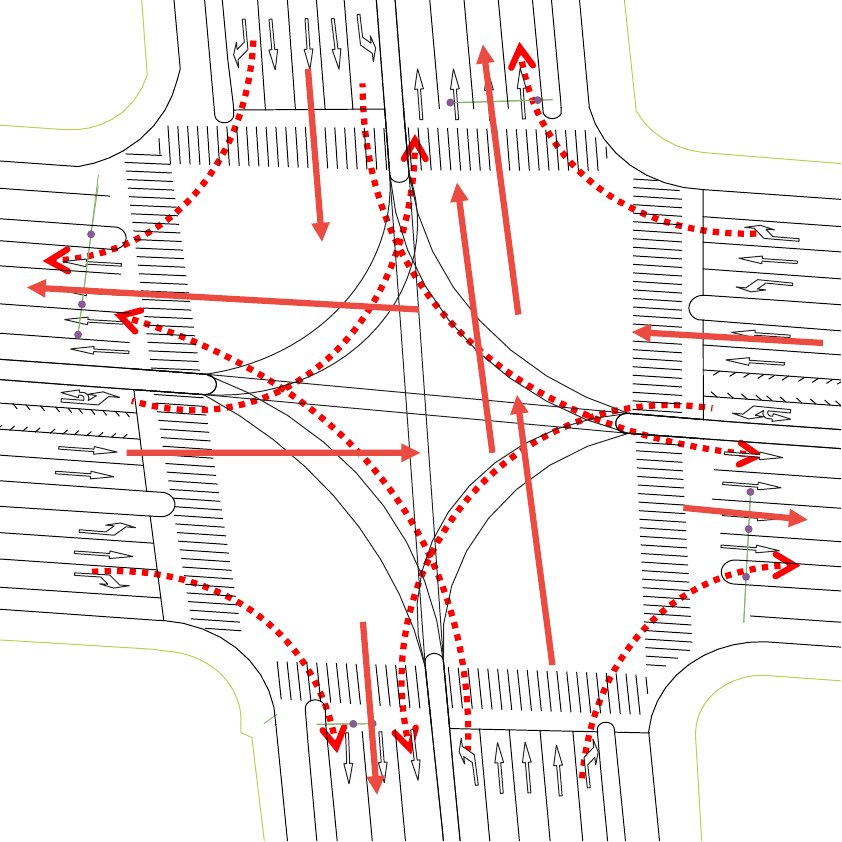}%
  \label{h1}}
  \hfil
  \subfloat[]{\includegraphics[width=0.33\columnwidth]{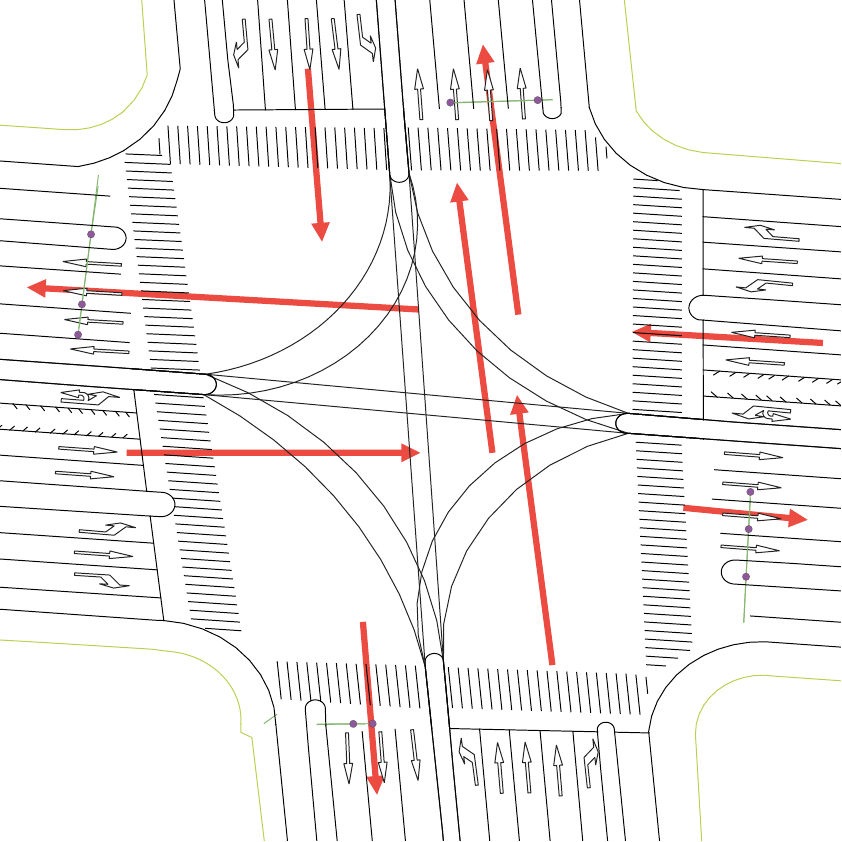}%
  \label{h2}}
  \subfloat[]{\includegraphics[width=0.33\columnwidth]{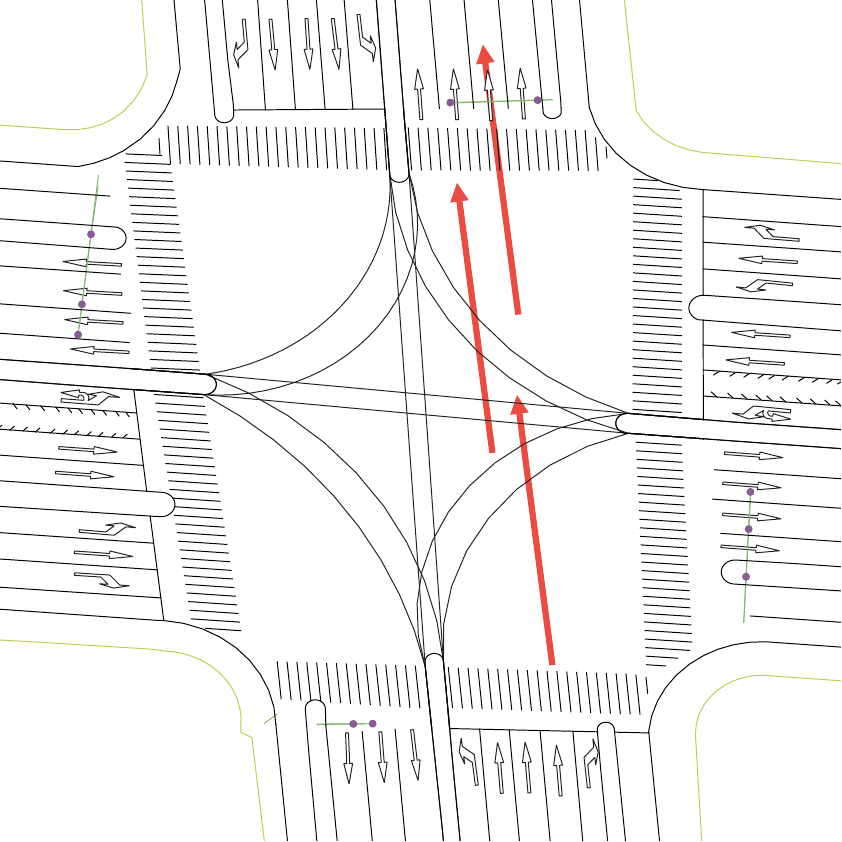}%
  \label{h3}}
  \caption{Illustration of Automatic Estimation of Camera Heading Angle. (a) Illustration of vehicle trajectory over a time interval. (b) Illustration of filtered vehicle trajectories, excluding non-straight ones and (c) Illustration of results after excluding trajectories based on a distance threshold.}
  \label{heading}
  \end{figure}
To ensure the Hungarian algorithm achieves high accuracy and establish initial monocular localization results for correlation with LiDAR-based positioning, we have developed a automatic camera heading angle estimation algorithm.
Traditional heading angle estimation can be directly measured through maps, but in large-scale scenarios, relying on manual methods is impractical. 
The algorithm is based on the majority of intersection scenarios where traffic cameras are oriented towards the direction of the lane. 
The core idea is to approximate the estimation of the camera heading angle $\omega_h$ through the estimation of the road heading angle $\omega_r$, which is presented in Algorithm \ref{algo:vehicle-selection-heading}.
Therefore, in this paper, we integrate vehicle GNSS data to approximate the automatic road heading angle estimation, thus adapting to a wide range of traffic scenarios. 

The Algorithm \ref{algo:vehicle-selection-heading} proceeds as follows: it starts by aggregating the trajectory geolocations of all vehicles within the time window.
Subsequently, it constructs trajectories as graphs, filtering out those that do not exhibit straight-line characteristics. 
Furthermore, in cases where a vehicle remains stationary for at least $T_t$ seconds, its distance $d$ from the camera is calculated. If this distance $d$ is less than the distance threshold $T_{h_d}$ between the vehicle and the camera, it is determined that the vehicle is oriented in the same direction as the camera heading angle $\omega_h$. 
In such instances, the vehicle is added to the heading angle calculation list for further analysis.
Finally, the algorithm computes heading angles from angle list based on initial and final geolocations, selecting the most frequent angle as $\omega_h$. 
Note that another advantage of such design is the implicit inclusion of the camera's yaw angle $\theta$ in both the camera heading angle $\omega_h$ and the road heading angle $\omega_r$, which are subsequently optimized as $\omega_h = \omega_{r} + \theta$.
\subsection{Vehicle Detection And Tracking}\label{sec:Vehicle Detection And Tracking}
In this study, YOLOX \cite{yolox2021} and Bytetrack \cite{Zhang2021ByteTrackMT} are employed for vehicle detection and tracking in surveillance cameras. 
Vehicle detection is performed in videos, while tracking and localization can be conducted in both video and high-definition map (HD map). 
Since our framework primarily focuses on intelligent vehicles, we exclusively detect vehicles and utilize the pixel coordinates of their 2D bounding box's bottom center point as inputs for subsequent geolocation estimation.

It should be noted that using the bottom center point of the 2D bounding box as an input for the localization algorithm introduces a certain margin of error. 
However, experimental findings show that this approach achieves sufficient localization accuracy for high-level objective matching and data fusion. 
Additionally, in specific traffic deployment scenarios, if accurate associations are established among multiple sensors, roadside cameras can leverage the 3D information from LiDAR. 
Therefore, we did not consider the 3D pixel positions of the vehicles in the image as inputs for localization.
\subsection{Geolocation Estimation}\label{sec:Angle Estimation}
After calibration, we obtain the pitch $\varphi$, roll $\psi$, and vfov $\alpha$ of the camera. 
In traffic scenes, the roll $\psi$ angle deviation is generally small compared to pitch $\theta$ angle, as a result, the roll $\psi$ angle has a minor impact on the final localization compared to the pitch $\varphi$ angle. 
Therefore, this study does not consider the effect of the roll angle.
\begin{figure}[!t]
  \centering
  \subfloat[]{\includegraphics[width=0.5\columnwidth]{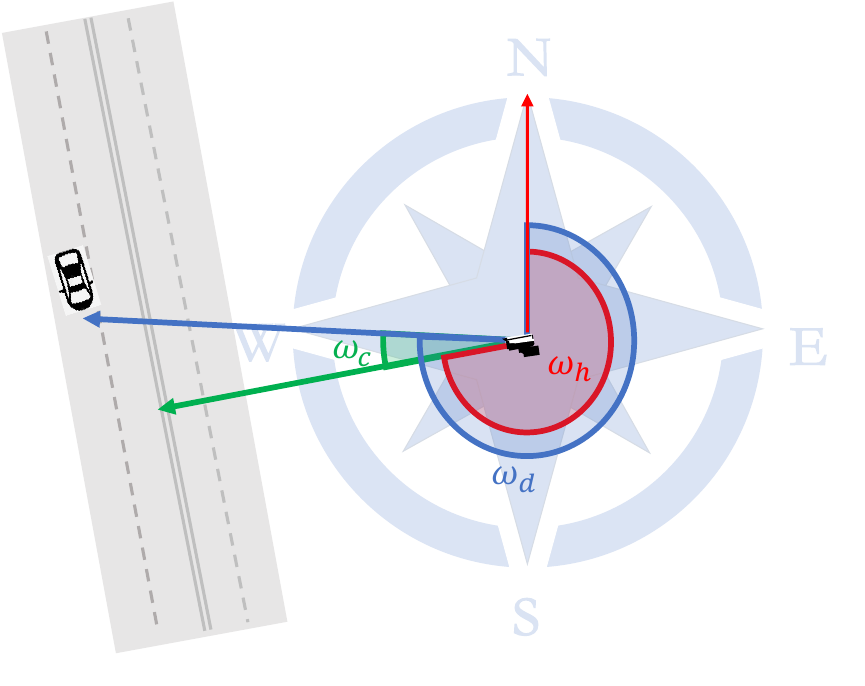}%
  \label{right_car}}
  \hfil
  \subfloat[]{\includegraphics[width=0.5\columnwidth]{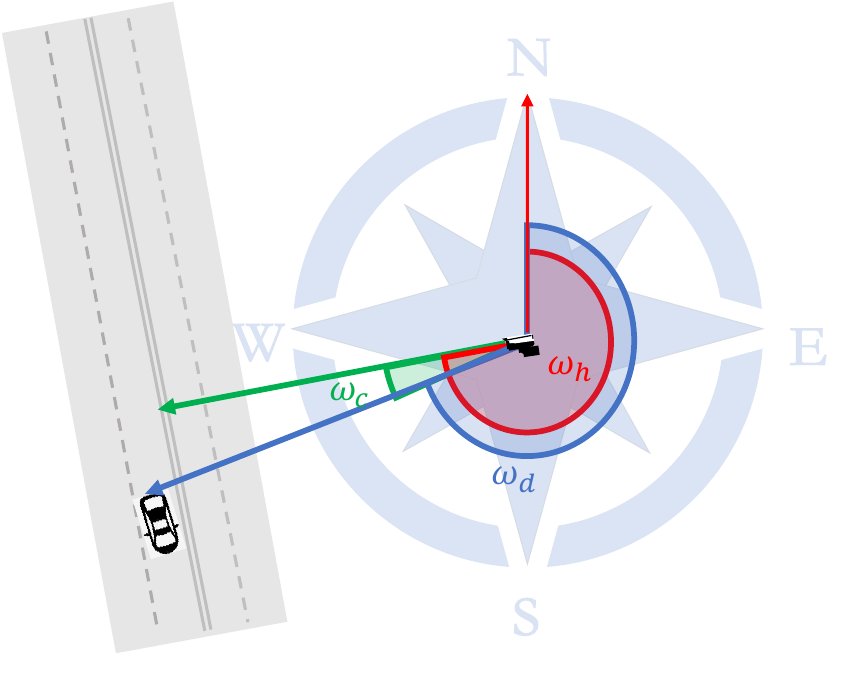}%
  \label{left_car}}
  \hfil
  \subfloat[]{\includegraphics[width=0.6\columnwidth]{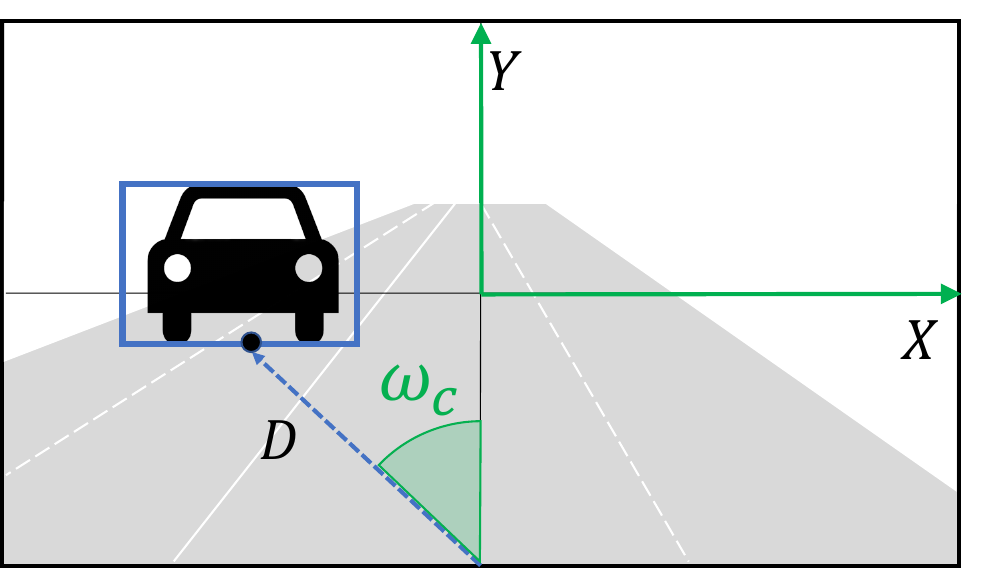}%
  \label{angle estimation}}
  \hfil
  \caption{The mathematical relations between $\omega_h$, $\omega_d$, and $\omega_c$, along with a comparison of (a) vehicle position on the right side, (b) vehicle position on the left side within the camera's field of view and (c) the image plane used in angle estimation.}
  \label{lr_angle_estimation}
  \end{figure}
  \begin{figure}[!t]
    \centering
    \includegraphics[width=\columnwidth]{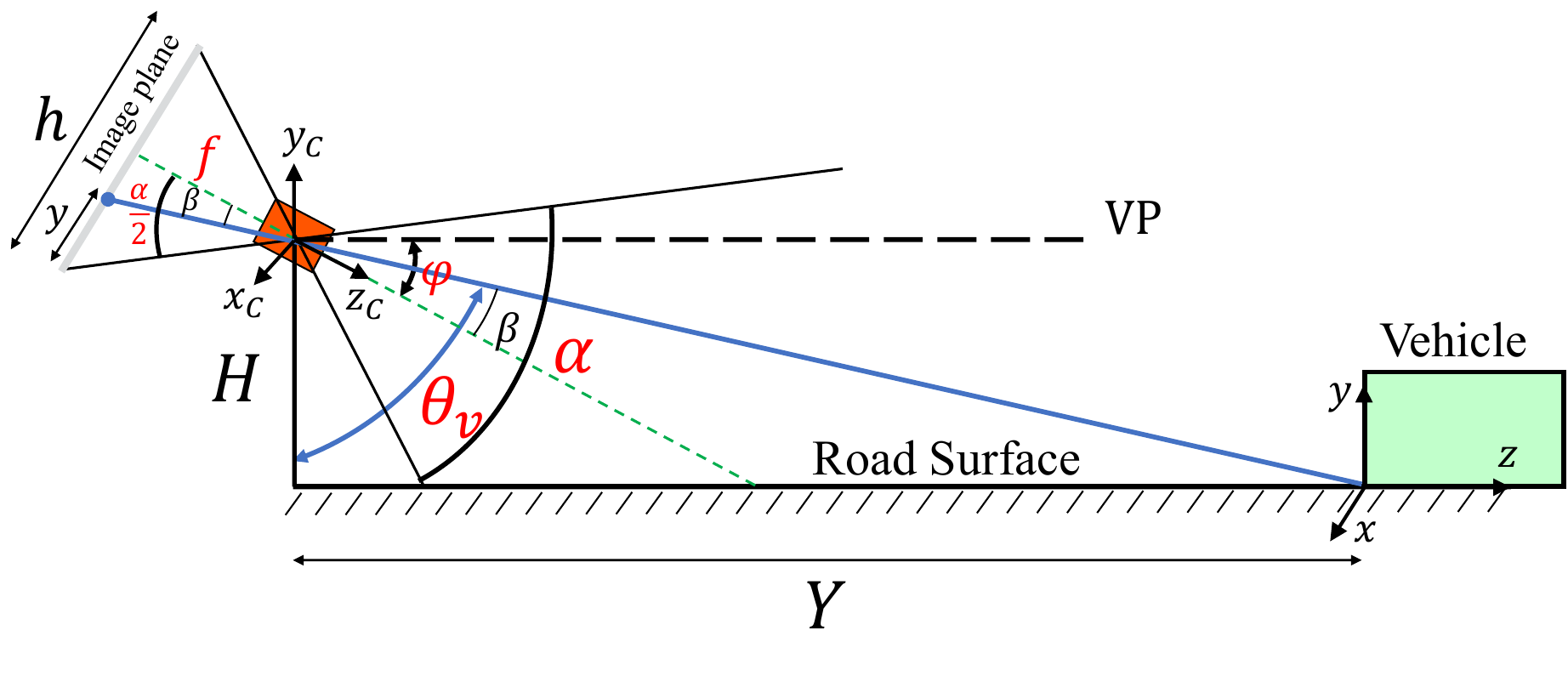}
    \caption{Real-world vehicle distance estimation based on pixel distance information in 2D image plane.}
    \label{fig:dis estimation}
  \end{figure}
As shown in the Fig. \ref{angle estimation}, given the position of the camera, we only need to know the distance $D$ and angle $\omega_d$ between the camera and the vehicle to estimate the geolocation of the target vehicle. 
In this paper, $\omega_d$ represents the clockwise angle between the distance $D$ and the north ($N$), and $\omega_c$ represents the angle between the vehicle and the camera heading angle $\omega_h$. 
Given an image width $w$ and camera $hfov$, the coordinates of the bottom center of the vehicle detection box are $(x, y)$, and $\omega_c$ can be expressed by: 
\begin{equation}
  \omega_c=\frac{hfov}{w}\left(x-\frac{w}{2}\right).
  \end{equation}
There are two different scenarios as shown in the Fig. \ref{lr_angle_estimation} \ref{sub@right_car} and Fig. \ref{lr_angle_estimation} \ref{sub@left_car}, where the vehicle's travel direction is located on the right and left side of the camera's field of view, respectively. In both cases, \eqref{eq:omega_c} remains accurate \cite{Namazi2022GeolocationEO}.
\begin{equation}
  \label{eq:omega_c}
  \omega_d\ =\ \omega_h\ +\ \omega_c.
  \end{equation}

The distance estimation model is depicted in Fig. \ref{fig:dis estimation}. Assume a detected vehicle in the road scene at an (unknown) position $(X_w,Y_w,Z_w)$, let $\theta_v$ be the angle of a projection ray (for the camera) pointing to the intersection of the planar approximation of the detected vehicle's rear or front portion with the planar road surface, $H$ is the mounting height of the camera \cite{rezaeiRobustVehicleDetection2015}. The longitudinal distance $Y$ between the camera and the vehicle can be calculated by:
\begin{equation}  Y=H \cdot\left[\tan \left(\frac{\pi}{2}-\theta+\tan ^{-1}\left(\frac{\frac{h}{2}-y}{f}\right)\right)\right].  \end{equation}
The distance $D$ is given by:
\begin{equation}
D = \frac{Y}{\cos\left(\omega_c\right)}.
\end{equation}
After obtaining the distance $D$ and angle $\omega_d$ between the vehicle and the camera, the latitude ($g_v$) and longitude ($l_v$) of the vehicle can be calculated using the following simplified formulas:
\begin{equation}
  \label{eq:latlon}
  \left\{
    \begin{aligned}
      &\quad arc = 6371.393 \times 1000,\\
      &\quad g_v = g_c + \frac{D \times \cos(\omega_d)}{\frac{\text{arc} \times 2 \times \pi}{360}},\\
      &\quad l_v = l_c + \frac{D \times \sin(\omega_d)}{\frac{\text{arc} \times \cos(g_v \times \pi / 180) \times 2 \times \pi}{360}},
    \end{aligned}
  \right.
  \end{equation}
where $g_c$ and $l_c$ represent the camera's latitude and longitude, $arc$ is the radius of the Earth in meters.
\section{SPATIAL SYNCHRONIZATION}\label{spatial synchronization}
\begin{figure}[b]
  \centering
  \subfloat[]{\includegraphics[trim=165 0 0 0, clip, width=0.53\columnwidth]{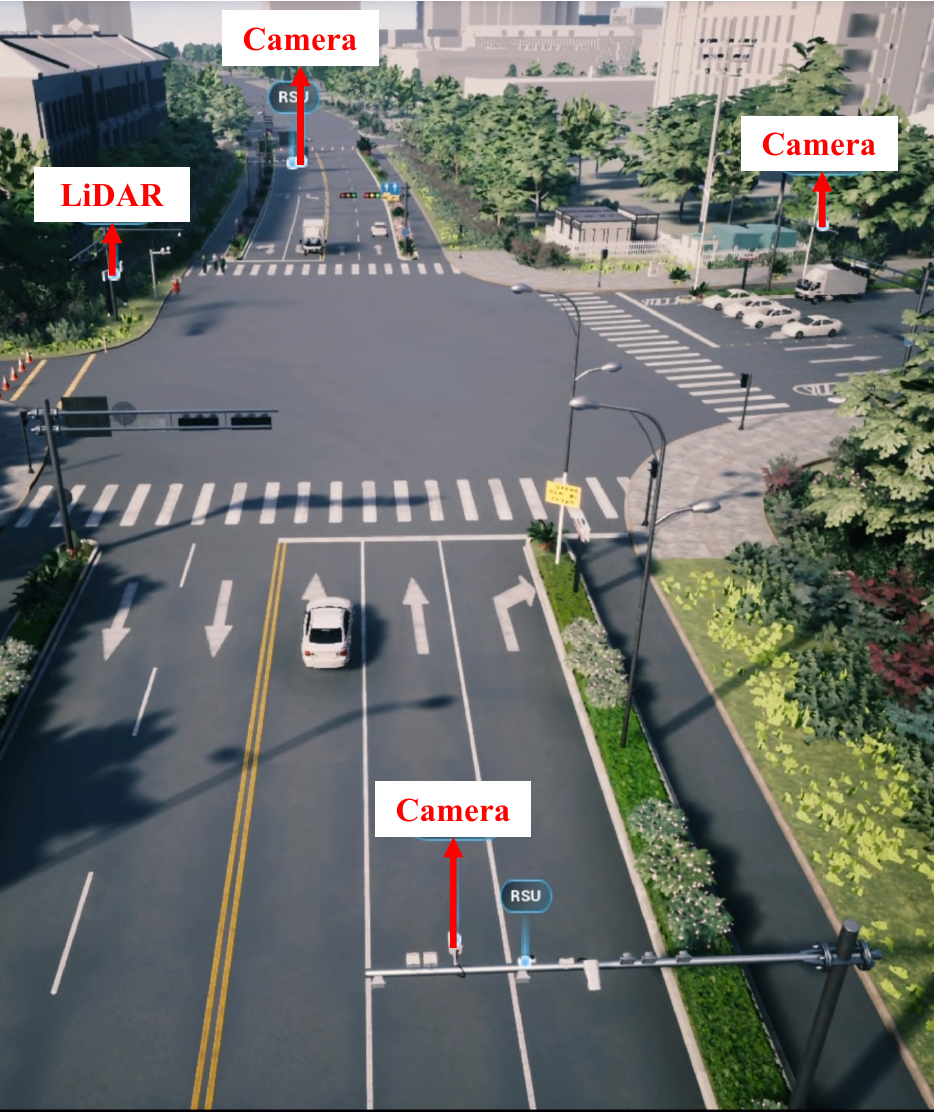}%
  \label{scene1}} 
  \hfil
  \subfloat[]{\includegraphics[trim=50 0 50 0, clip, width=0.47\columnwidth]{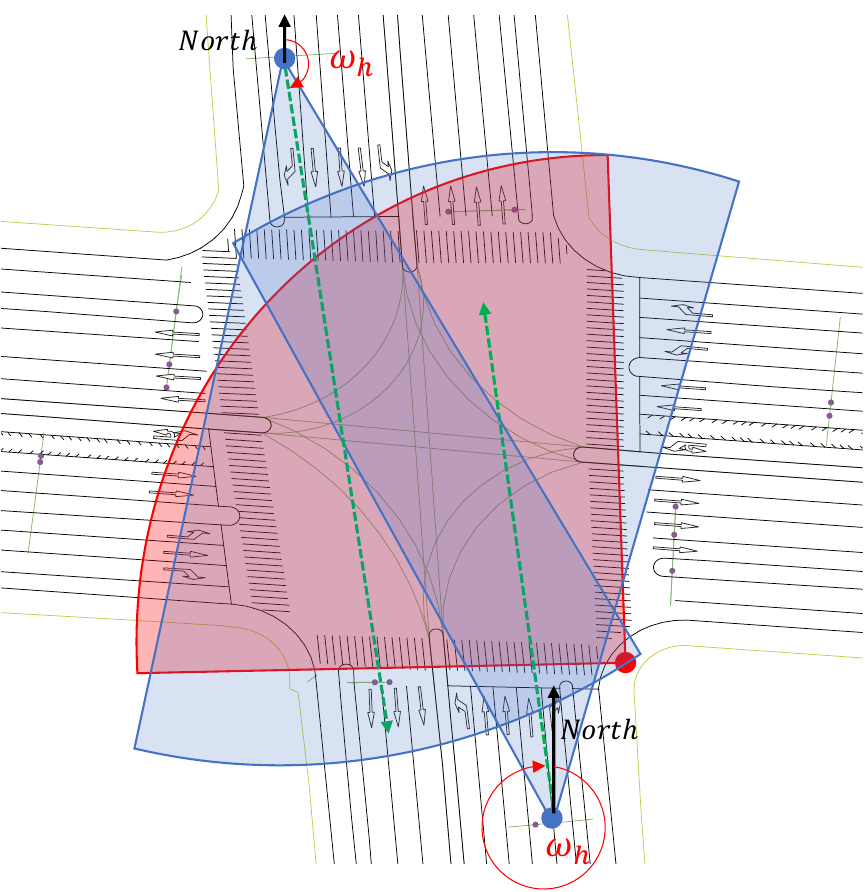}%
  \label{scene2}}
  \hfil
  \label{frame scene}
  \caption{(a) The schematic diagram of the scenario, and (b) the fusion area of the cameras and the LiDAR.}
  \end{figure}
Most roadside monocular localization methods \cite{luCAROMVehicleLocalization2021, Lu2022RealTimePL} rely on manually selecting at least four corresponding points between the image and the map, solving for the homography matrix, and achieving the conversion from pixel coordinates to world coordinates. 
However, this manual point selection process may introduce localization errors in practical scenarios.
For the non-manual point selection method proposed in this paper, using pre-trained camera calibration models from public datasets and predicted camera parameters for localization calculations may fall short in achieving spatial synchronization in roadside scenarios.
Because accurate global parameters of the camera, such as latitude $g_c$, longitude $l_c$, height $H$, and heading angle $\omega_h$, are also necessary in addition to the focal length $f$ and pitch angle $\varphi$ as mentioned earlier in Section \ref{problem statement and framework}.
Obtaining precise values for these parameters in large-scale traffic scenarios is extremely challenging.

To overcome this practical application challenge, our approach involves treating these parameters as a unified entity, denoted as $\phi_i = \{ f, \varphi, H, \omega_h \}$.
Through the utilization of the vehicle's LiDAR geolocation data, we optimize monocular localization accuracy while concurrently fine-tuning the values within $\phi_i$.
\subsection{Outlier Tracking IDs Filtering}\label{sec:Objective Filtering}
It is crucial to preclude vehicles outside the fusion area before utilizing the Hungarian algorithm \cite{kuhnHungarianMethodAssignment1955, Munkres1957AlgorithmsFT}. 
This procedure can enhance the initial association accuracy by mitigating three critical practical challenges that lead to inaccuracies when directly employing the Hungarian algorithm for association:
\begin{enumerate}
  \item{Different intersections exhibit diverse sensor attributes and perception algorithms (e.g., object detection algorithms) with varying levels of precision.}
  \item{The Camera and LiDAR installation positions, mounting heights, and orientations can significantly differ.}
  \item{Dense vehicle scenarios inevitably introduce occlusions, ID switches, and tracking interruptions, affecting the output of each sensor.}
\end{enumerate}
\begin{algorithm}[!t]
  \caption{Outlier Rejection for Tracking Ids}
  \label{algo:outlier-rejection}
  \begin{minipage}[t]{0.8\textwidth}
    \textbf{Input:} $\omega_h$, $T_{a}$, $T_{d}$, $T_{v}$, $T_{p}$, $(f_i, d_i, v_i)$ for $i \in \{1, 2, \ldots, N_{l}\}$ \\
    and $\bm{p_j}$ for $j \in \{1, 2, \ldots, N_{c}\}$ \\
    \textbf{Output:} Filtered tracking id list $L^*, C^*$ in LiDAR and camera
  \end{minipage}
  \begin{algorithmic}[1]
    \STATE \textbf{Initialize:} filtered LiDAR data list $L^*$
    \STATE \textbf{Initialize:} filtered pixel data list $C^*$
    \FOR{$i$ from 1 to $N_{l}$} 
        \IF{$|mean(f_i) - \omega_h| > T_{a}$}
            \STATE \textbf{continue} \hfill $\triangleright$ Skip vehicles outside fusion area or moving towards the camera
        \ENDIF
        
        \IF{$min(d_i) > T_{d}$}
            \STATE \textbf{continue} \hfill $\triangleright$ Skip vehicles too far from camera
        \ENDIF
        
        \IF{$max(v_i) < T_{v}$}
            \STATE \textbf{continue} \hfill $\triangleright$ Skip stationary vehicles
        \ENDIF
        
        \STATE $L^* \leftarrow L^* \cup \{i\}$ \hfill $\triangleright$ Add LiDAR tracking id to $L^*$
    \ENDFOR
    \FOR{$j$ from 1 to $N_{c}$} 
        \IF{$min(\bm{p}_{v_j}) < T_{p}$}
            \STATE \textbf{continue} \hfill $\triangleright$ Skip vehicles too far from camera
        \ENDIF
        \STATE $\Delta \theta_{\bm{p}_j} = PixelAverageDisplacementCal(\bm{p}_{v_j})$
        \IF{$\Delta \theta_{\bm{p}_j} < 0$ or $\Delta \theta_{\bm{p}_j} \approx 0$}
            \STATE \textbf{continue} \hfill $\triangleright$ Remove vehicles moving towards the camera or stationary
        \ENDIF
        
        \STATE $C^* \leftarrow C^* \cup \{j\}$ \hfill $\triangleright$ Add camera tracking id to $C^*$
    \ENDFOR
    \RETURN $L^*, C^*$
  \end{algorithmic}
\end{algorithm}
These three challenges result in variations in both the quantity and characteristics of vehicles detected by each sensor, thereby increasing errors in association algorithms. 
Moreover, an excessive quantity of initial matches may result in prolonged association times, thereby impacting the computational efficiency of the association algorithm.

Before performing the association, it is necessary to extract a time window. 
In this paper, the time window is defined as the vehicle's start-up time, determined using Algorithm \ref{algo:vehicle-selection-heading} or extracted directly from traffic signal interface.
This choice is based on the relatively low initial monocular localization error during this period, which contributes to improved association stability.
The Algorithm \ref{algo:outlier-rejection} relies on LiDAR tracking ID $L_i = (f_i, d_i, v_i)$, encompassing the orientation, distance, and face of each vehicle, as well as the tracking results from the video: $\bm{p_j} = (u_j, v_j)$.
The LiDAR tracking outliers are filtered based on distance threshold $T_{d}$, angle threshold $T_{a}$, and speed threshold $T_{v}$ to exclude vehicles that are either too far from the camera, outside the fusion area, or stationary. 
Similarly, video tracking outliers are filtered using pixel threshold $T_{p}$ and inter-frame relative displacement $\Delta \theta$ to exclude vehicles outside the fusion area and stationary ones.
\subsection{Multi-object Association}
In this paper, the data association is initialized using the Kuhn-Munkres algorithm \cite{kuhnHungarianMethodAssignment1955, Munkres1957AlgorithmsFT}. 
The correct data association is a prerequisite to ensure the convergence of the loss during the optimization of camera parameters, and the primary objective of this step is to generate a small dataset for subsequent training purposes. 
Typically, the data association is carried out once, assuming reliable associations. 
Nevertheless, in cases where the optimization of camera parameters faces difficulties in loss convergence due to incorrect associations, the association is performed repeatedly until the loss converges successfully, ensuring the accuracy of the camera parameter optimization. 

After preprocessing the camera and LiDAR data, we obtain less noisy datasets denoted as $L^*$ and $C^*$. 
These datasets respectively contain the tracking results from the LiDAR and camera sensors, which are used for subsequent weight computation. 
The specific procedure involves starting with the initial camera parameter set $\phi_i$ to derive the initial vehicle localization results. 
These results encompass latitude and longitude coordinates denoted as $g_j$, the angle $\omega_{d_j}$ between the vehicle and the camera, and the distance $d_j$ between them. 
The corresponding results from the LiDAR sensor are represented as $g_i$, $\omega_{d_i}$, and $d_i$. 
The weights between these two sets of data are computed based on the values from the first frame, following the weight calculation formula below:
\begin{equation}
  \label{eq:weight}
  \mathbf{W}_{lc} = \sum_{i=1}^{N_{l^*}} \sum_{j=1}^{N_{c^*}} \left( k_1 \cdot |d_i - d_j| + k_2 \cdot |\omega_{d_i} - \omega_{d_j}| \right),
\end{equation}
where $N_{l^*}$ and $N_{c^*}$ denote the counts of tracked IDs in the filtered LiDAR and camera data, respectively, following the removal of outliers.
After obtaining the weights, we apply the Hungarian algorithm directly on the weight matrix to determine the vehicle matching results. 
\subsection{Camera Parameters Approaching}
Since the localization algorithm in this paper is differentiable, the losses between the predicted distances \( \hat{d}_i \) and angles \( \hat{a}_i \) from the model and the ground truth \( d_i \) and \( a_i \) obtained from the LiDAR can be calculated, and the camera parameters can be continuously updated to enhance the localization accuracy by utilizing backpropagation \cite{Rumelhart1986LearningRB} and stochastic gradient descent.
As in \eqref{eq:h, f}, there exists a deterministic relationship between the focal $f$ and the vfov $\alpha$. 
Therefore, during the optimization process, we only compute gradients for the $\alpha$, which is then used to update the values of the angles $f$ and $hfov$. 
Given $n$ samples, geolocation loss $\mathcal{L}_{loc}$ is determined by the following formula:
\begin{equation}
  \mathcal{L}_{loc} = \frac{1}{n} \sum_{i=1}^{n} \left( (\hat{a}_i - a_i)^2 + (\hat{d}_i - d_i)^2 \right).
\end{equation}

We minimize using stochastic gradient descent with the Adam optimizer \cite{Kingma2014AdamAM}, employing an initial learning rate of $\eta$ = 0.001. 
The algorithm's efficiency contributes to its rapid training and its ability to perform well with limited data.
When the best parameter combination for the $i$-th camera, denoted as $\phi_i^*$, is obtained, and the pixel position of a vehicle within the common detection area in the $i$-th camera is given by $\bm{p_i} = (u_i, v_i)$, its corresponding geolocation $\bm{q_i}$ can be represented as $\bm{q_i} = (l_i, g_i)$. 
If there are a total of $N$ cameras deployed in the scene, the parameter combination for the multi-view cameras system is denoted as $\Phi^* = \{ \phi_1^*,\phi_2^* \dots \phi_N^* \}$. 
The overall process of multi-view spatial synchronization can then be expressed as follows:
\begin{equation}
  \left( \phi_1^*,\phi_2^* \dots \phi_N^* \right) \begin{pmatrix}
    p_{1}\\
    p_{2} \\
    \vdots  \\
    p_{n}
    \end{pmatrix}=\begin{pmatrix}
    q_{1} \\
    q_{2} \\
    \vdots  \\
    q_{n}
    \end{pmatrix}.
  \end{equation}
  \begin{figure}[!b]
    \centering
    \includegraphics[width=\columnwidth]{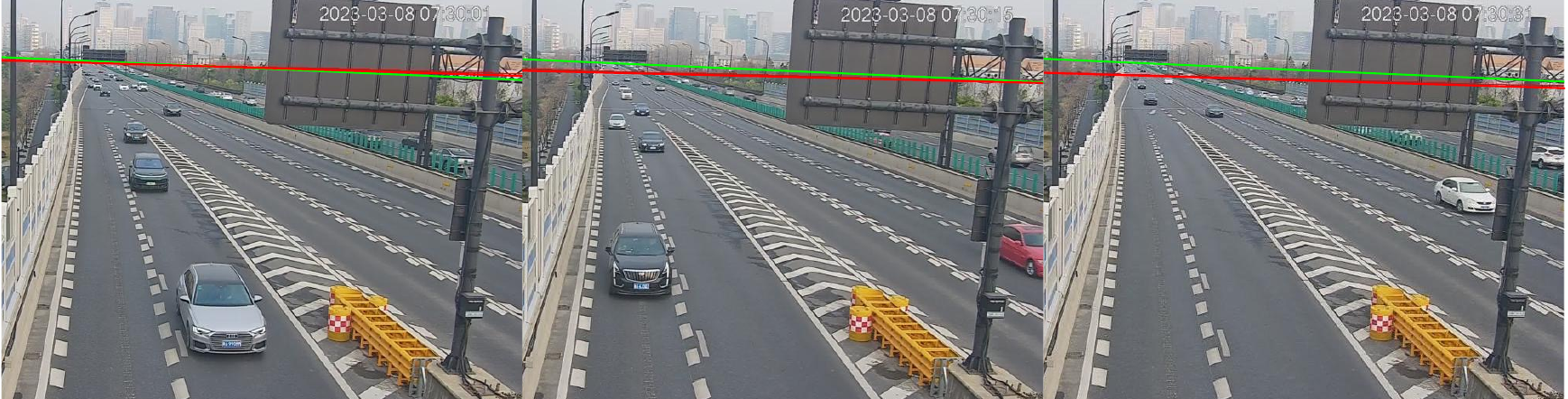}
    \caption{The impact of different traffic flows on the accuracy of camera calibration. The green and red lines represent the true and predicted horizontal lines respectively.}
    \label{camera calibration}
    \end{figure}
\section{EXPERIMENTAL RESULTS}\label{experimental results}
Our experiment data is collected from two intersections and one highway in Hangzhou, China. 
\subsection{Localization Evaluation}
\begin{figure}[!t]
  \centering
  \subfloat[]{\includegraphics[width=0.5\columnwidth]{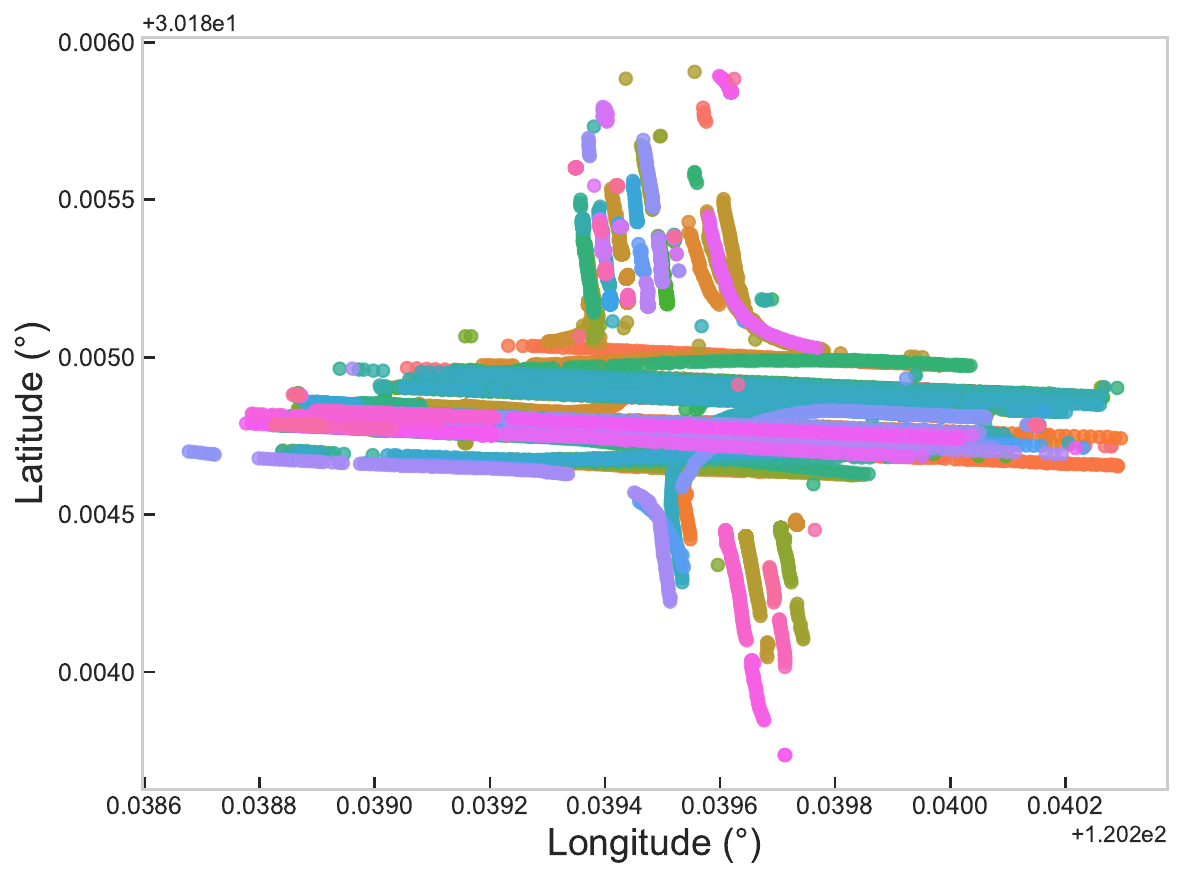}%
  \label{heading_choose_1}}
  \hfil
  \subfloat[]{\includegraphics[width=0.5\columnwidth]{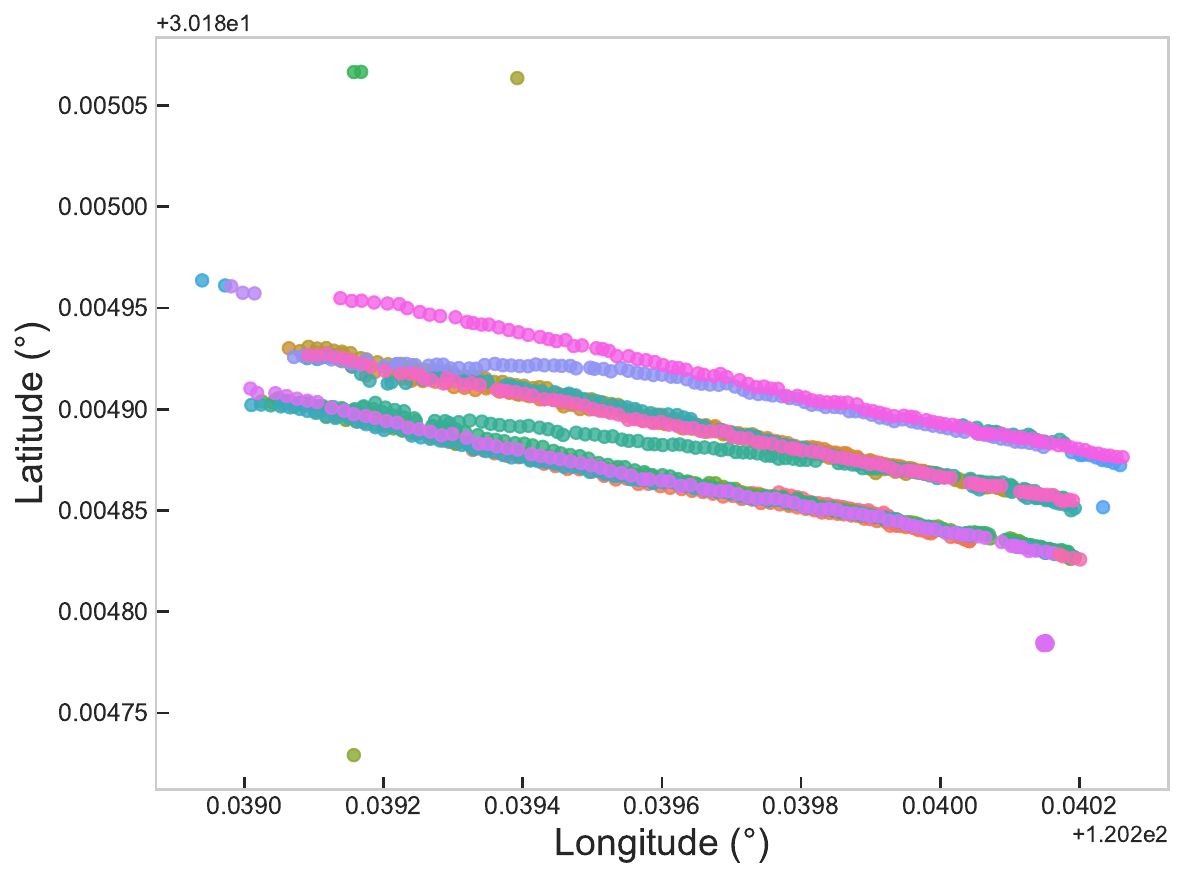}%
  \label{heading_choose_3}}
  \hfil
  \subfloat[]{\includegraphics[width=0.5\columnwidth]{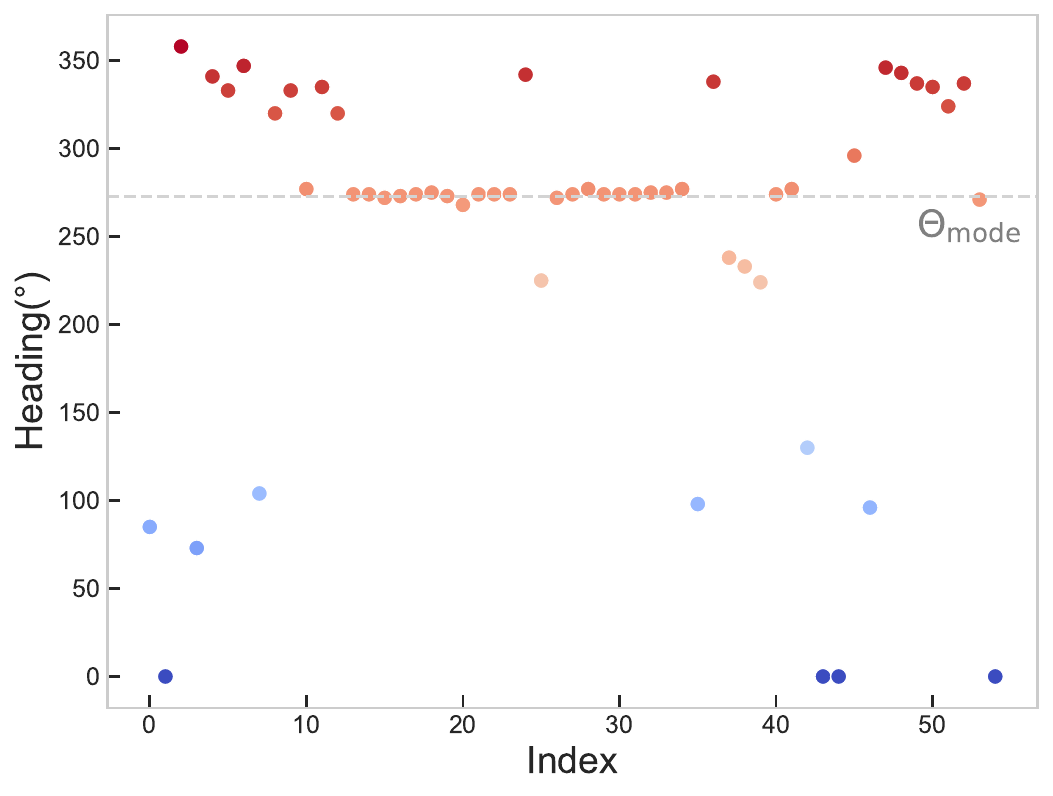}%
  \label{heading_estimation}}
  \hfil
  \subfloat[]{\includegraphics[width=0.4\columnwidth]{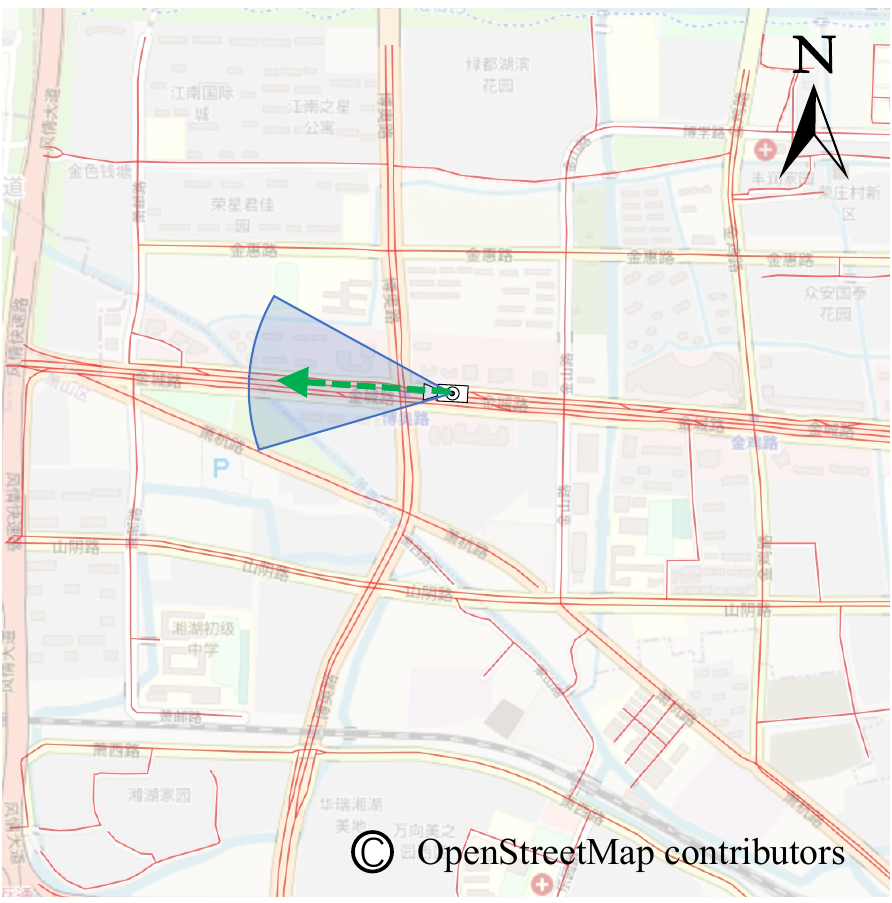}%
  \label{map}}
  \caption{Automatic camera heading estimation results: (a) the LiDAR vehicle GNSS trajectory over a certain period, (b) the remaining trajectories after applying Algorithm \ref{algo:vehicle-selection-heading}, filtering out only the straight-moving vehicles aligned with the camera's heading $\omega_h$, (c) the $\Theta_{\text{mode}}$ estimated by Algorithm \ref{algo:vehicle-selection-heading}, (d) the map of the corresponding scene, map data copyrighted OpenStreetMap contributors and available from https://www.openstreetmap.org \cite{OpenStreetMapContributors2017}.}
  \label{heading estimation}
\end{figure}

The Pano360 dataset \cite{kocabasSPECSeeingPeople2021} was used for training camera calibration model. Fig. \ref{camera calibration} reveals the impact of different traffic flows on the accuracy of camera calibration. It was observed that our camera calibration model is affected to some extent by the presence of traffic, but the resulting errors remained within an acceptable range.
As depicted in Fig. \ref{heading estimation}, the camera heading angle $\omega_h$ was obtained through Algorithm \ref{algo:vehicle-selection-heading}, utilizing a camera situated on the east side of the intersection (facing west). 
In this paper, $T_t$ and $T_{h_d}$ in Algorithm \ref{algo:vehicle-selection-heading} are set to 5 s and 30 m, respectively.
The results in Fig. \ref{heading estimation}\ref{sub@heading_estimation} showed that $\omega_h$ closely aligned with the orientation $\omega_r$ of the roads in Fig. \ref{heading estimation}\ref{sub@map}, which was derived from the map provided by OpenStreetMap\cite{OpenStreetMapContributors2017}.

\begin{figure}[!t]
  \centering
  \subfloat[]{\includegraphics[width=0.5\columnwidth]{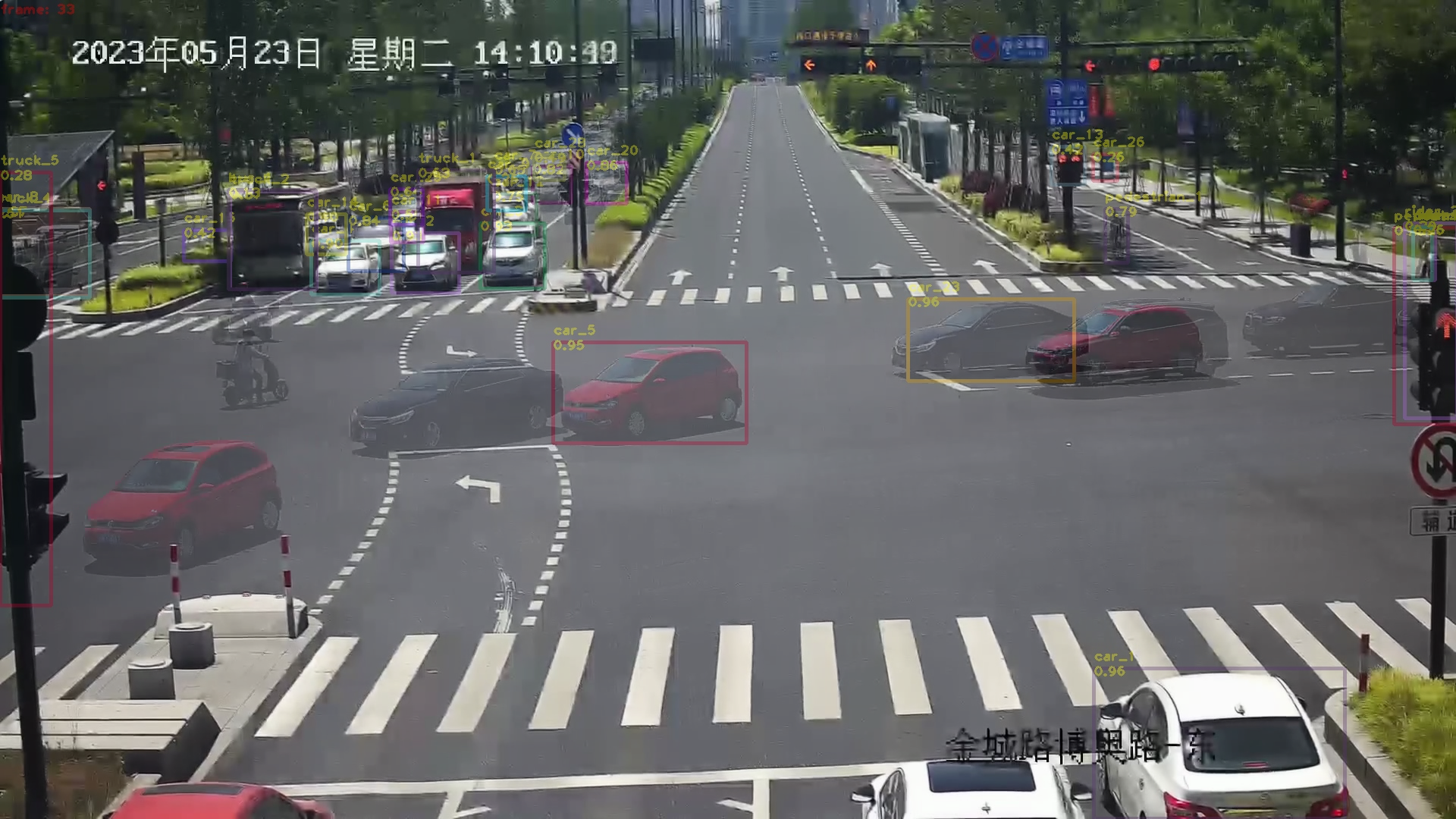}%
  \label{stright}}
  \hfil
  \subfloat[]{\includegraphics[width=0.5\columnwidth]{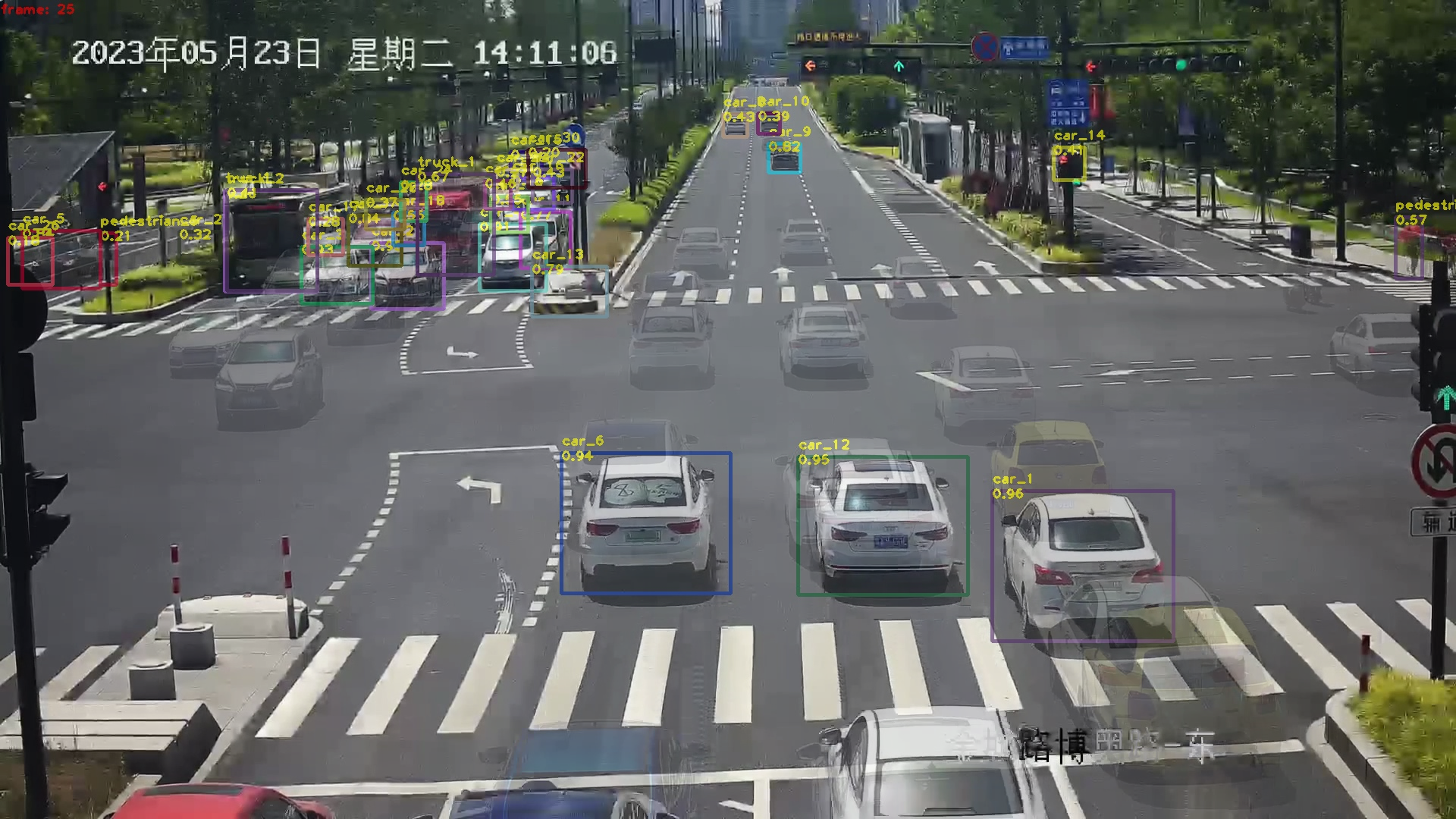}%
  \label{turn}}
  \hfil
  \subfloat[]{\includegraphics[width=0.5\columnwidth]{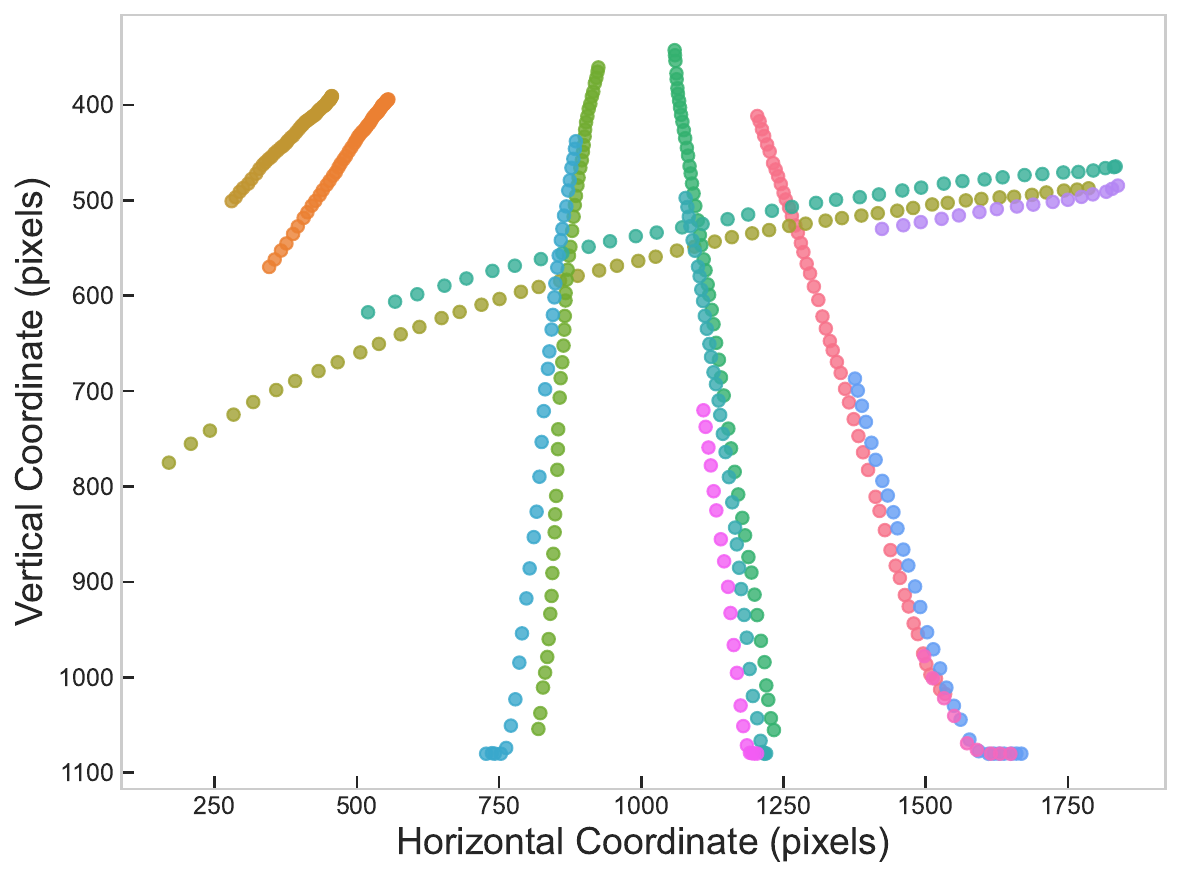}%
  \label{pixel_location}}
  \hfil
  \subfloat[]{\includegraphics[width=0.5\columnwidth]{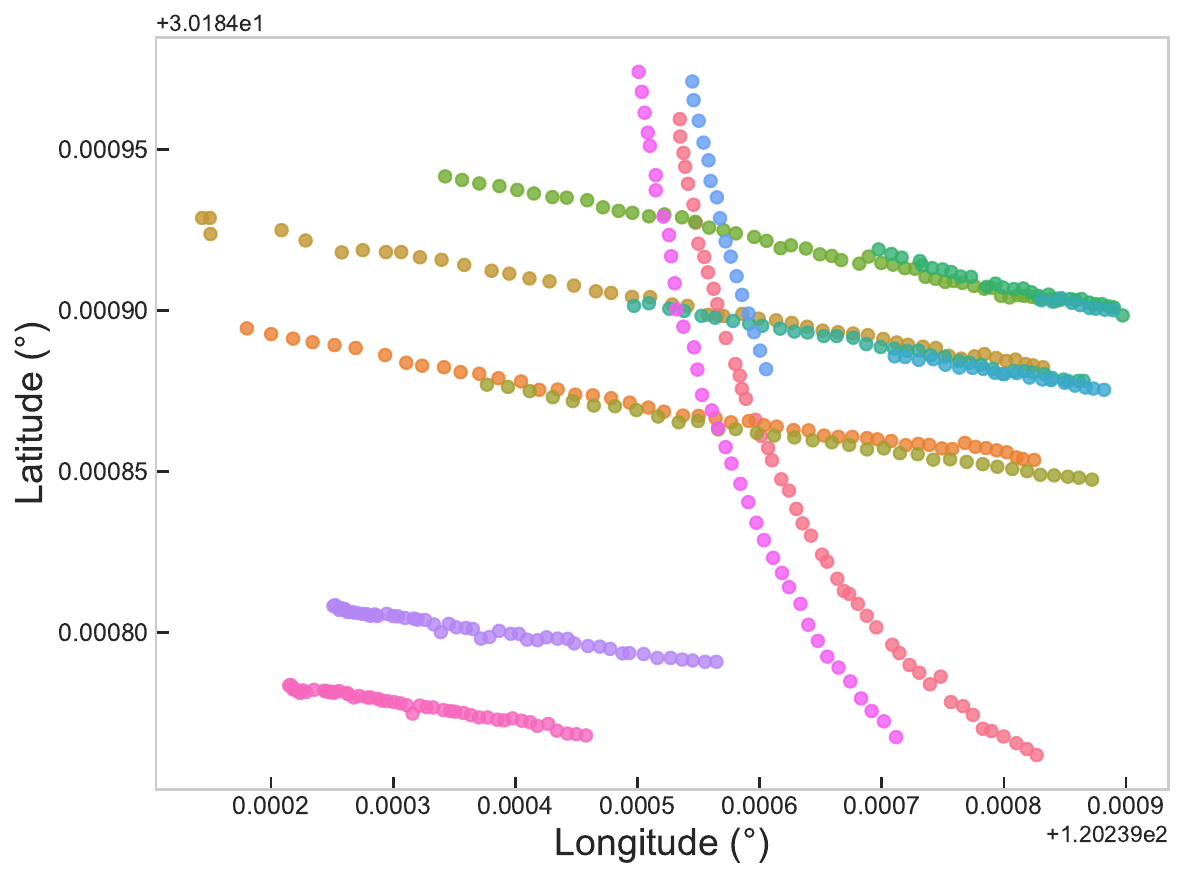}%
  \label{lidar_localization}}
  \hfil
  \caption{Experimental Selection of Vehicle Trajectories: (a) Turning Scenario, (b) Straight-line Scenario, (c) Pixel Variations, (d) LiDAR GNSS Trajectories.}
  \label{scenes}
\end{figure}

\begin{figure}[!t]
  \centering
  \subfloat[]{\includegraphics[width=0.5\columnwidth]{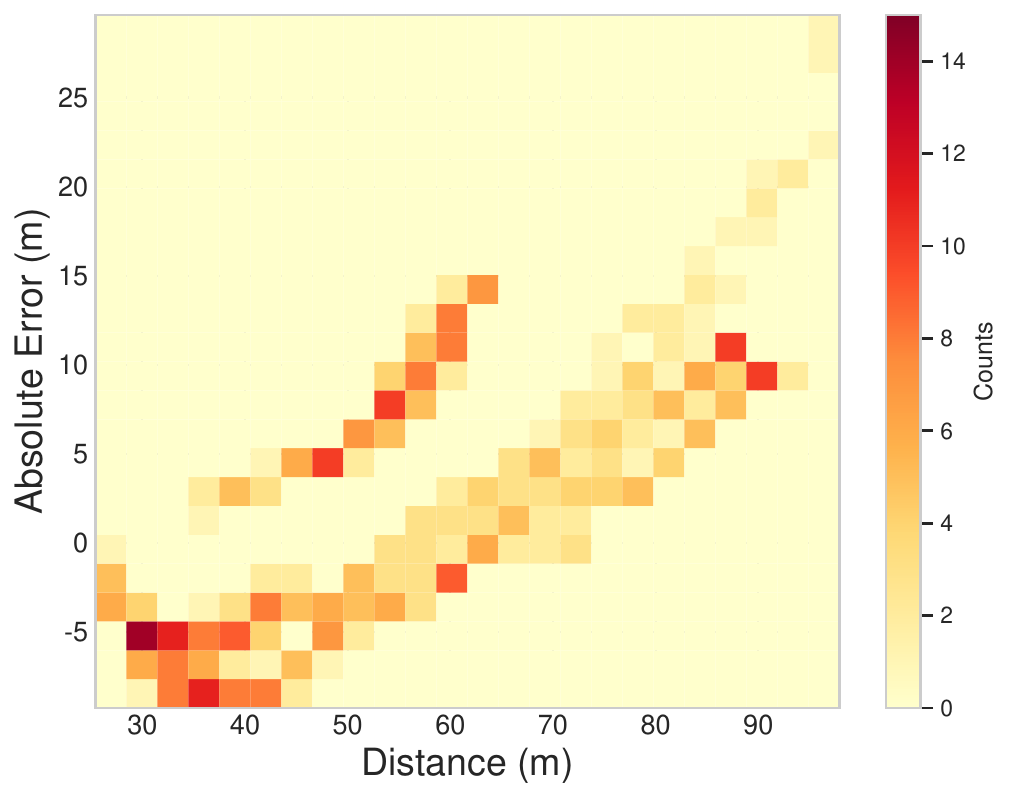}%
  \label{all_car_dis}}
  \hfil
  \subfloat[]{\includegraphics[width=0.5\columnwidth]{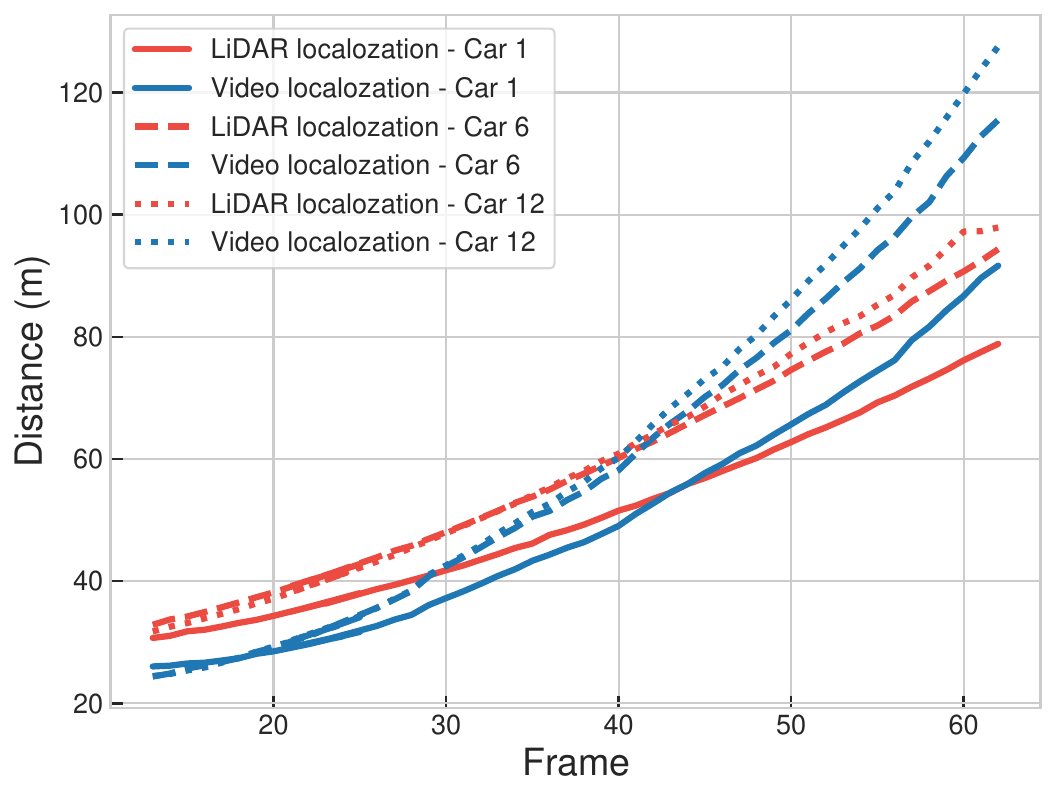}%
  \label{three_car_dis}}
  \hfil
  \subfloat[]{\includegraphics[width=0.5\columnwidth]{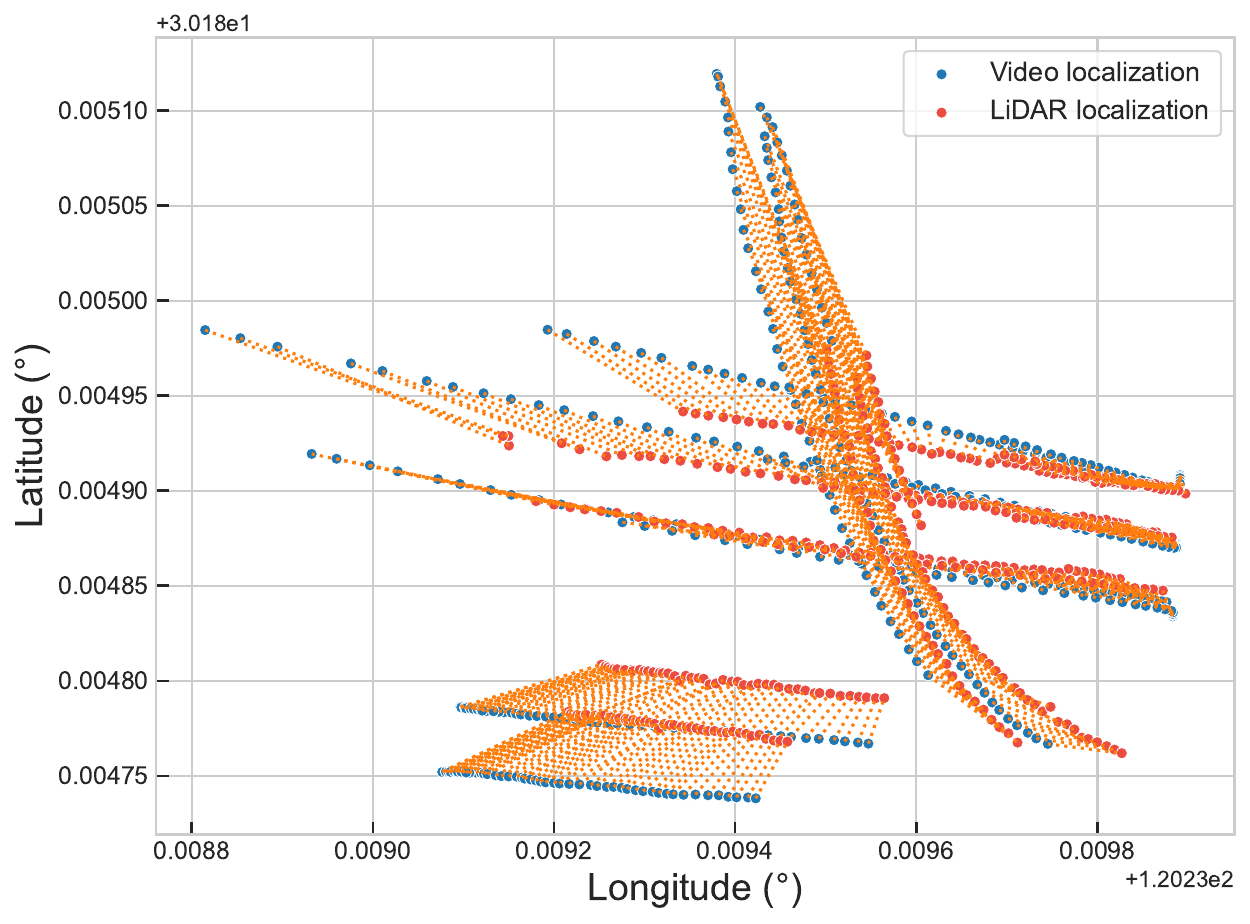}%
  \label{raw_cam_param}}
  \hfil
  \subfloat[]{\includegraphics[width=0.5\columnwidth]{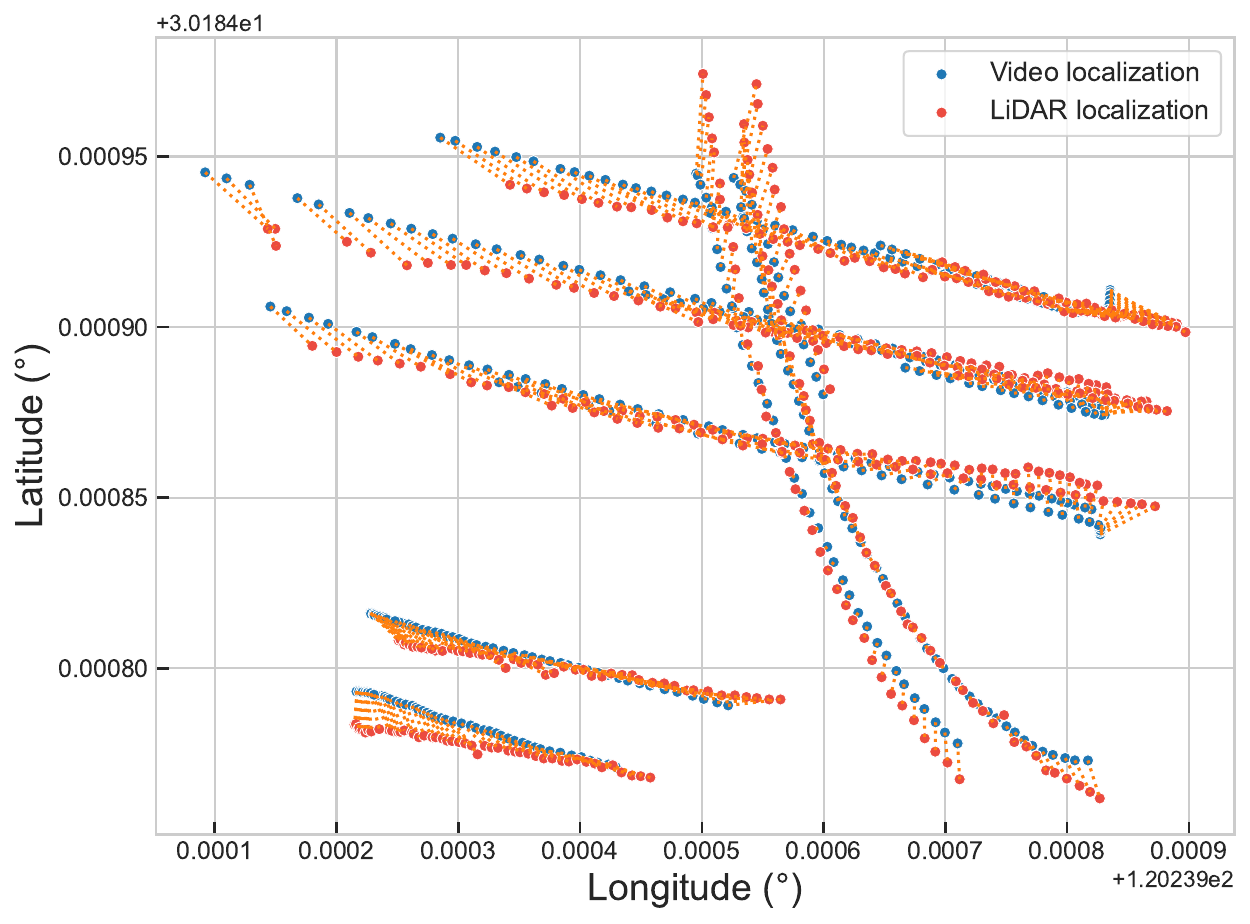}%
  \label{new_cam_param}}
  \caption{Distance Error in Monocular Localization and Analyzing Localization Results through Vehicle Trajectory Overlays in Scenarios: (a) Heatmap of Distance Error Distribution, (b) Frame-wise Analysis for Three Vehicles. (c) The trajectory Before Camera Parameter Optimization, (d) The trajectory After Camera Parameter Optimization}
  \label{car dis and location}
\end{figure}

We conducted experiments in two primary scenarios as shown in Fig. \ref{scenes} and \ref{scenes}\ref{sub@stright}, namely, both sampled over a 5-second duration. Given the disparity in sampling frequencies between LiDAR and video data, we standardized them to 10fps, assuming time synchronization. The turning scenario, as depicted in Fig. \ref{scenes}\ref{sub@turn}, involved three vehicles. In the straight-line phase, as seen in Fig. \ref{scenes}\ref{sub@stright}, we selected eight vehicles moving westward (in the same direction as the $\omega_h$) and two vehicles moving eastward (opposite to $\omega_h$).
Localization experiments were conducted using the camera calibration parameters obtained in the previous step, along with the camera's heading angle from Algorithm \ref{algo:vehicle-selection-heading}, initial camera mounting height $H$, and camera latitude $g_c$ and longitude $l_c$. 
The results of the experimental error distribution are presented in Fig. \ref{car dis and location}\ref{sub@all_car_dis}, and individual frame localization results are shown in Fig. \ref{car dis and location}\ref{sub@three_car_dis}. 
Results from Fig. \ref{car dis and location}\ref{sub@all_car_dis} indicate that absolute errors are concentrated in the range of 5 to 10 m. 
It is evident that using the initial parameters $\phi$ from the outset for localization results in significant errors. 
We believe that the factors contributing to these errors include, but are not limited to, errors in the localization algorithm itself, errors in camera parameters (obtained through calibration algorithms, initial height $H$, camera heading angle $\omega_h$, and camera geolocation), and time asynchrony. 
Additionally, from Fig. \ref{car dis and location}\ref{sub@all_car_dis}, \ref{sub@three_car_dis} and \ref{sub@raw_cam_param}, it can be observed that as the distance increases, the absolute distance error also increases. Different distance distributions exhibit varying errors, primarily attributable to different angles. Compared to straight-line vehicles, turning vehicles exhibit larger angle errors and distance errors.
\subsection{Approximating Evaluation}
  \begin{figure*}[!t]
    \centering
    \subfloat[]{\includegraphics[width=0.5\columnwidth]{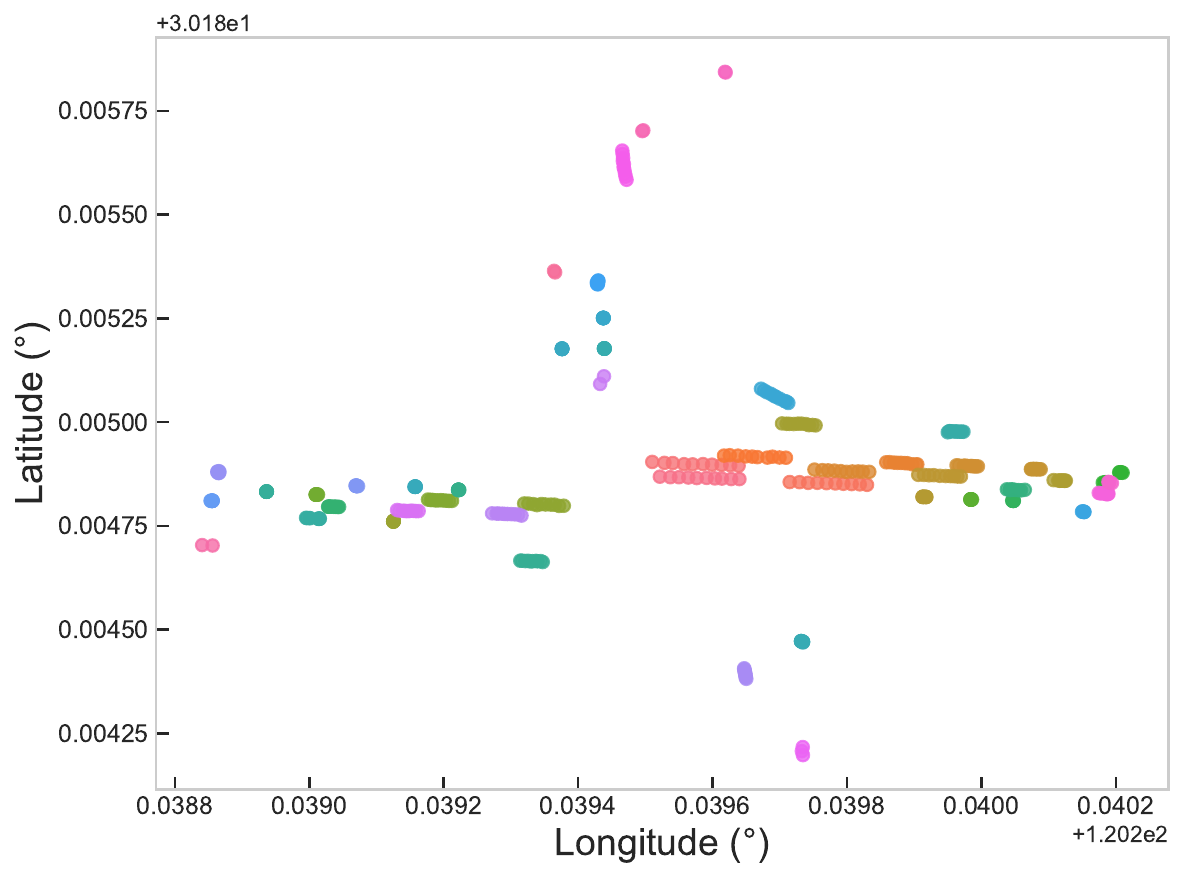}%
    \label{filtering1_l}}
    \hfil
    \subfloat[]{\includegraphics[width=0.5\columnwidth]{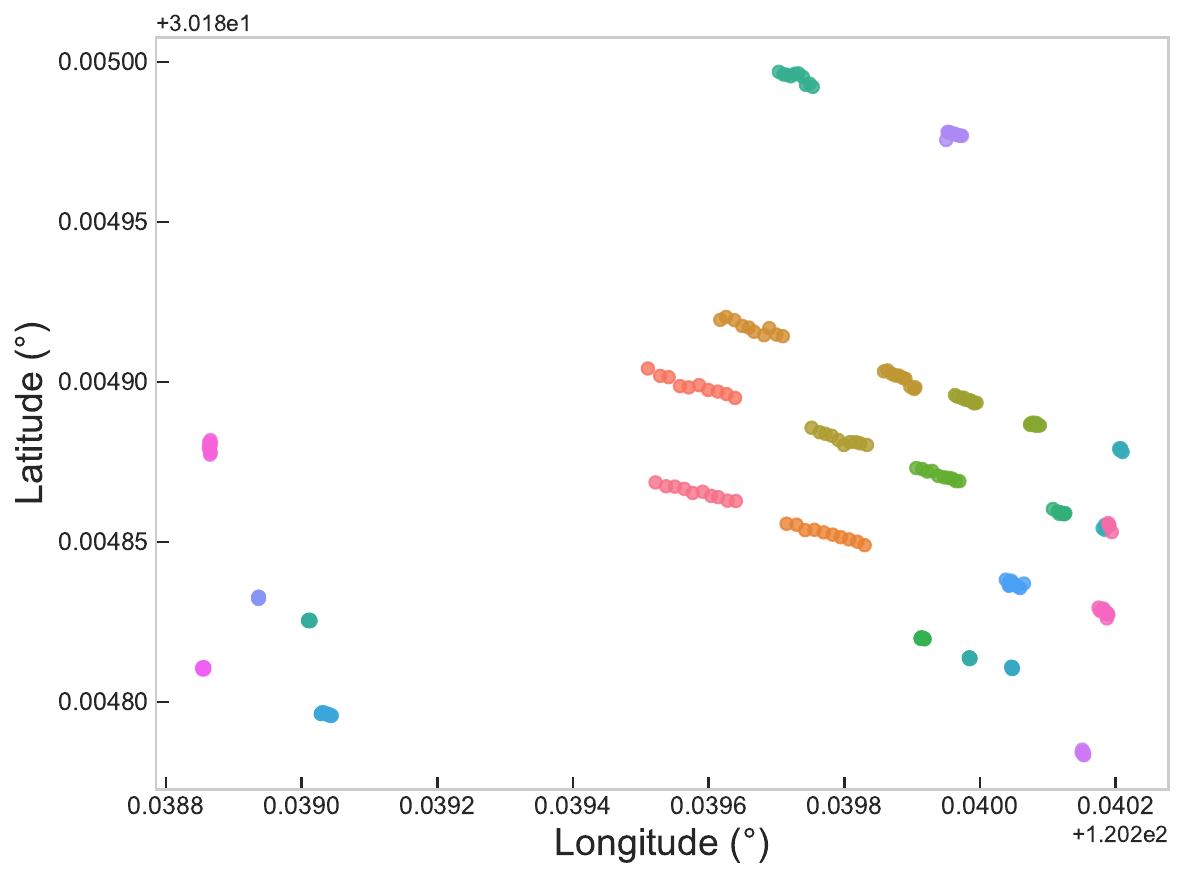}%
    \label{filtering2_l}}
    \hfil
    \subfloat[]{\includegraphics[width=0.5\columnwidth]{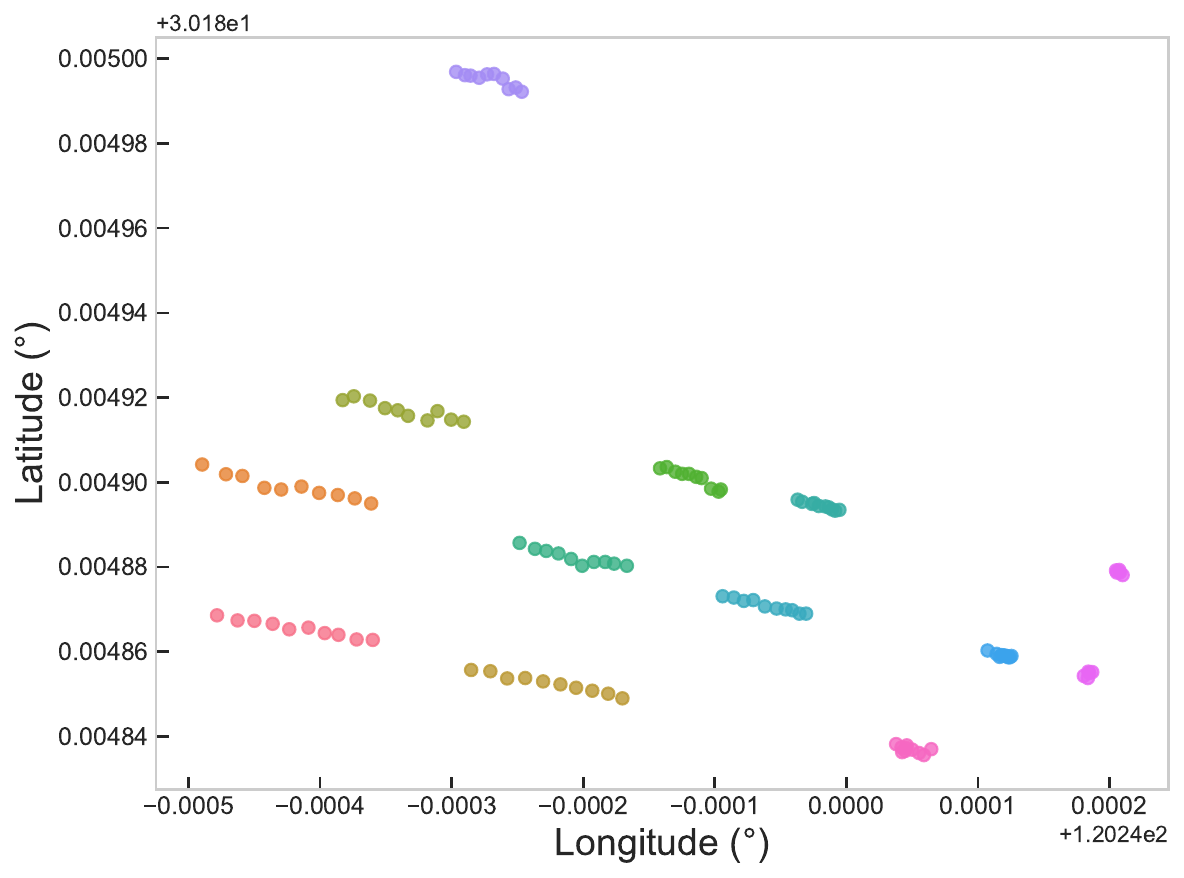}%
    \label{filtering3_l}}
    \hfil
    \subfloat[]{\includegraphics[width=0.5\columnwidth]{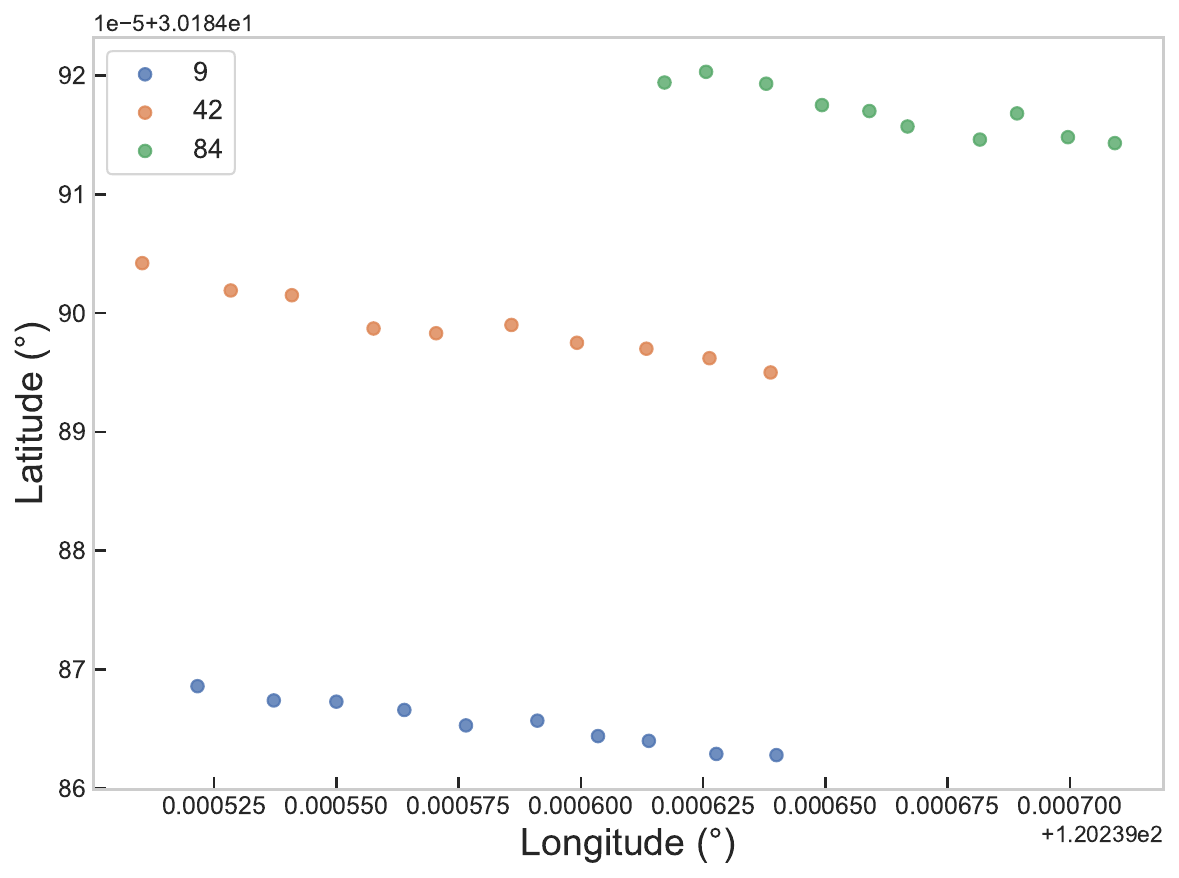}%
    \label{filtering4_l}}
    \hfil
    \subfloat[]{\includegraphics[width=0.5\columnwidth]{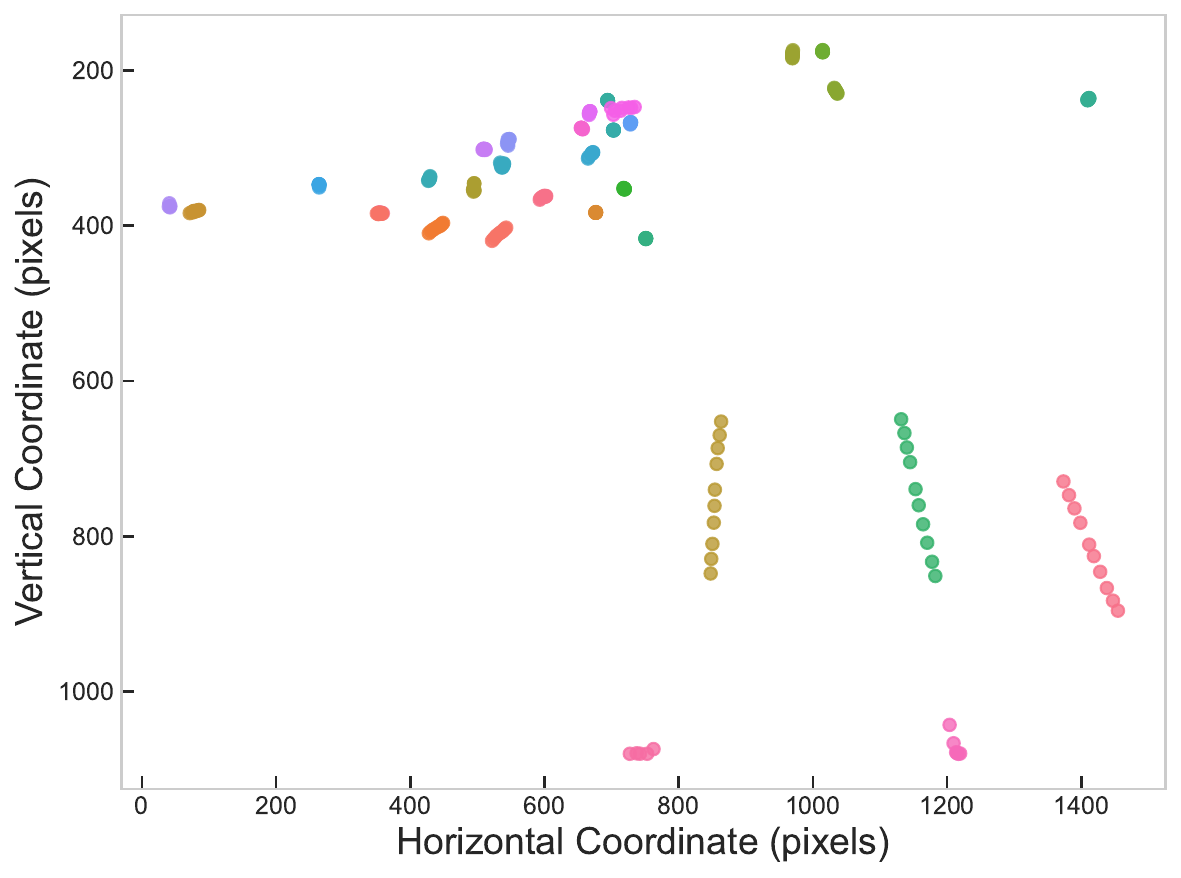}%
    \label{filtering1_v}}
    \hfil
    \subfloat[]{\includegraphics[width=0.5\columnwidth]{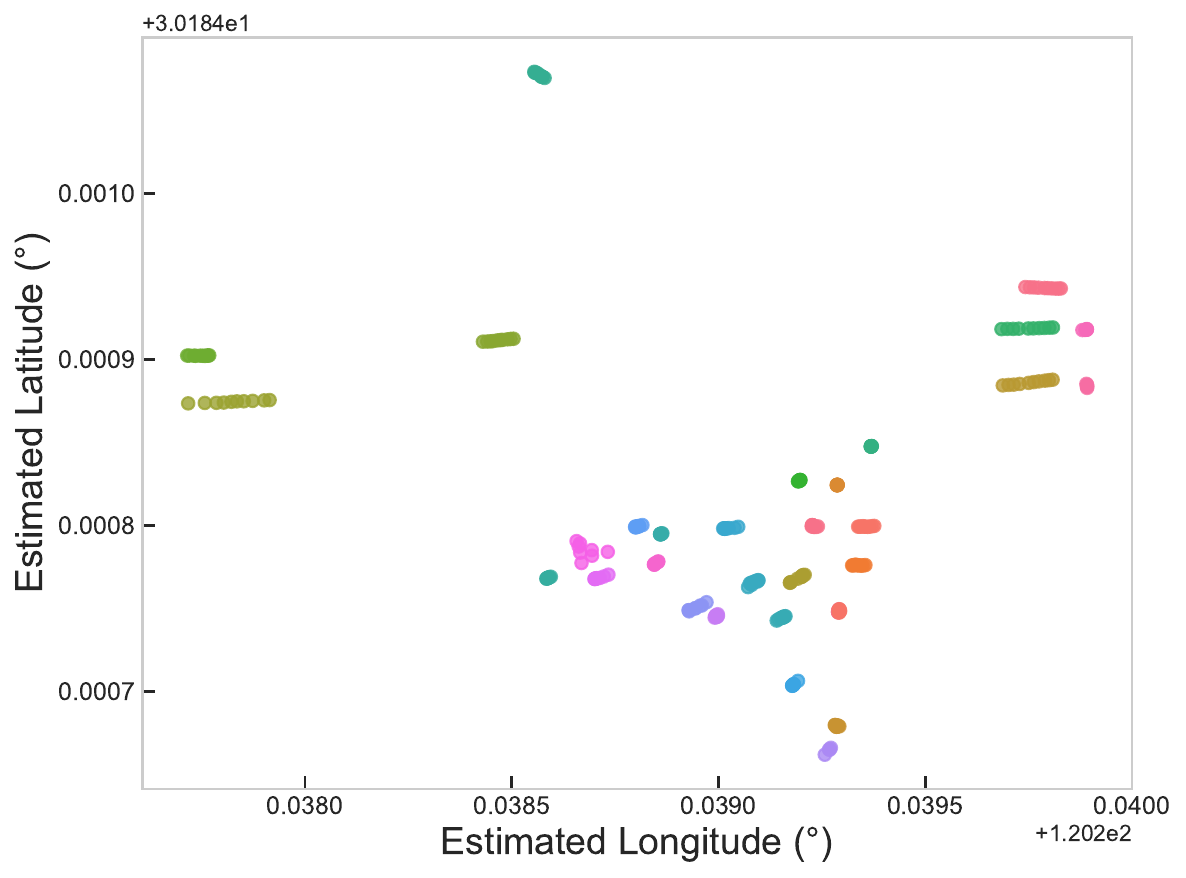}%
    \label{filtering2_v}}
    \hfil
    \subfloat[]{\includegraphics[width=0.5\columnwidth]{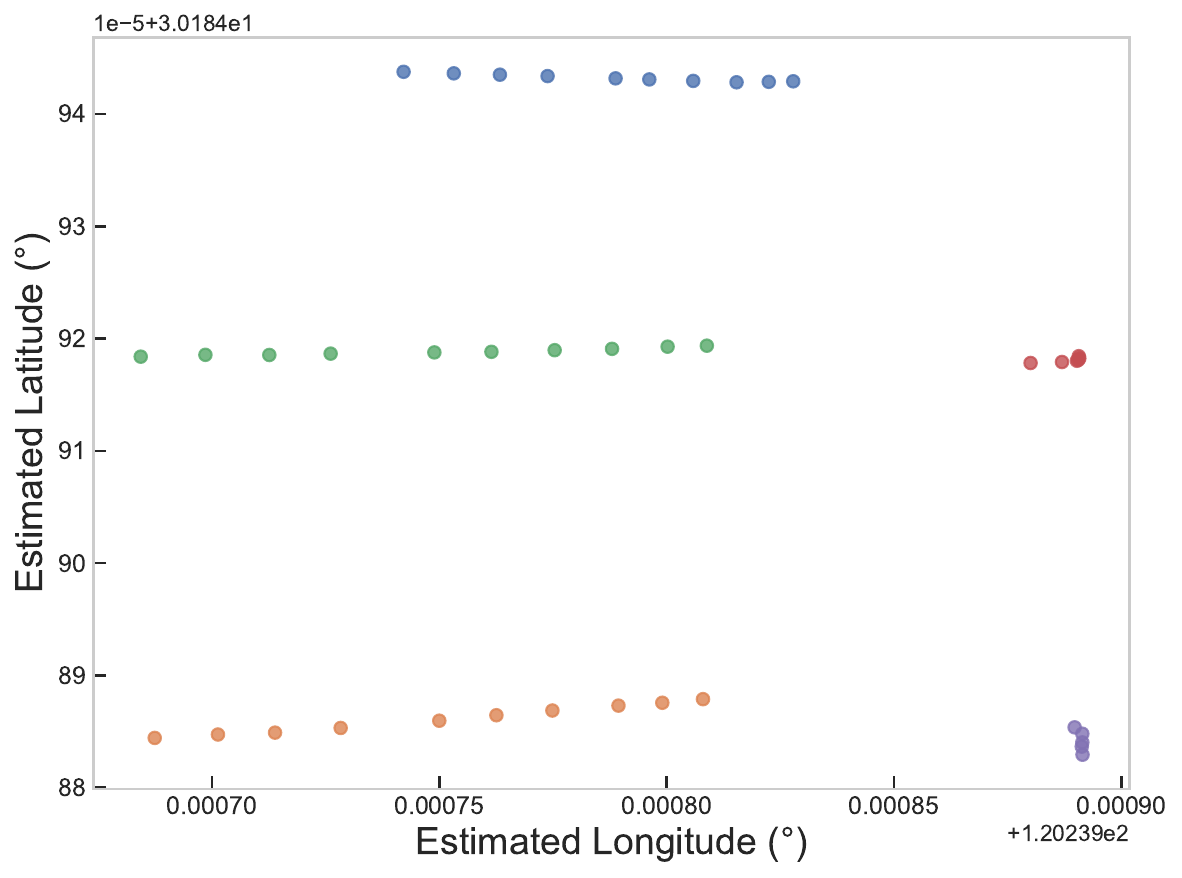}%
    \label{filtering3_v}}
    \hfil
    \subfloat[]{\includegraphics[width=0.5\columnwidth]{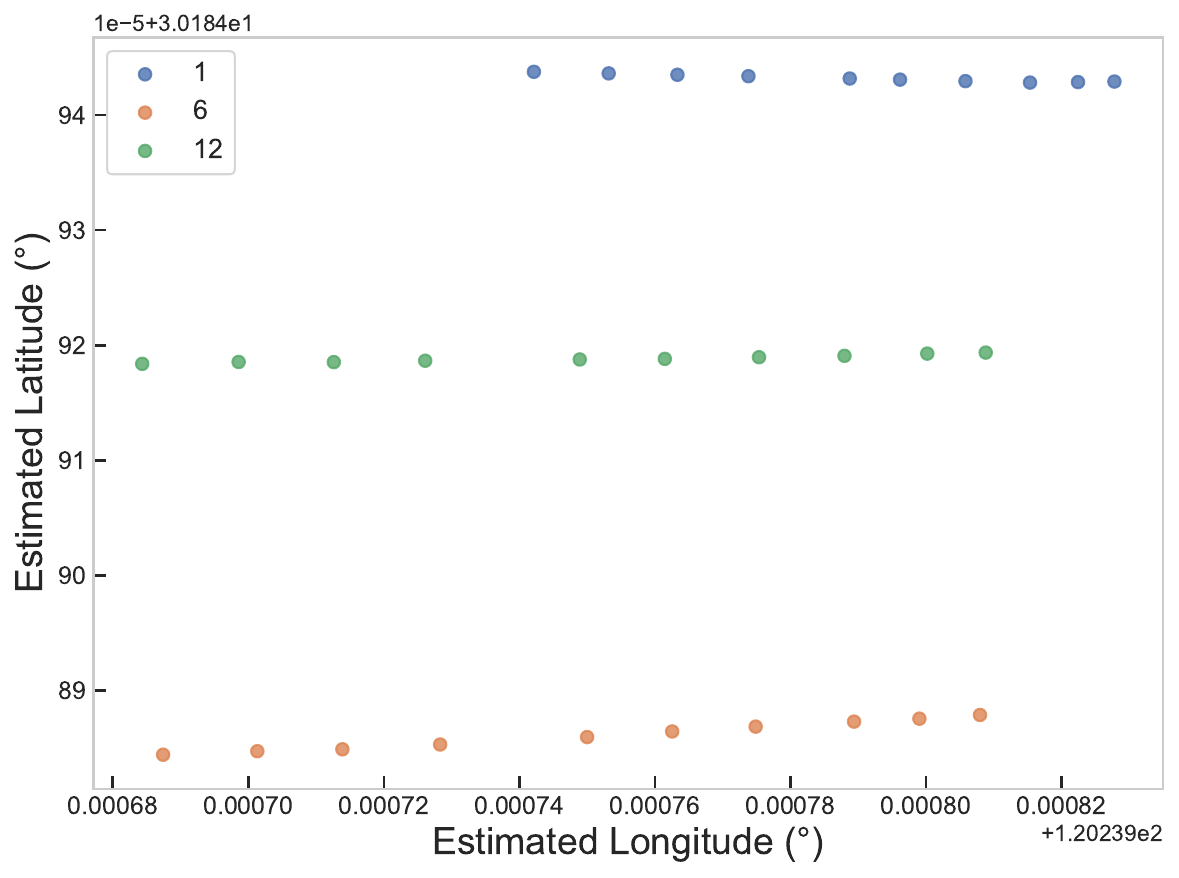}%
    \label{filtering4_v}}
    \caption{Analyzing Localization Results through Vehicle Trajectory Overlays in Three Scenarios: (a) Before Camera Parameter Optimization, (b) After Camera Parameter Optimization}
    \label{lidar video outlier}
  \end{figure*}
  \begin{table}[!t]
    \caption{THE PARTIAL STRUCTURED DATA (ONE VEHICLE) OF VIDEO AND MMW LiDAR DETECTION.}
    \label{tab:YourTableLabel}
    \centering
    \begin{tabular}{cccccc}
      \toprule
      \multicolumn{1}{c}{\multirow{2}{*}{\textbf{Frame}}} & \multicolumn{2}{c}{\textbf{Video localization}} & \multicolumn{2}{c}{\textbf{LiDAR Localization}} \\
      \cline{2-5}
       & \textbf{Distance (m)} & \textbf{Angle (\textdegree)} & \textbf{Distance (m)} & \textbf{Angle (\textdegree)} \\
      \toprule
      1 & 60.59 & 275.87 & 60.61 & 273.30 \\
      2 & 60.50 & 275.31 & 60.34 & 272.96 \\
      3 & 60.19 & 274.79 & 60.10 & 272.57 \\
      4 & 60.02 & 274.35 & 59.70 & 272.06 \\
      5 & 59.76 & 273.80 & 59.27 & 271.63 \\
      6 & 59.28 & 273.13 & 58.89 & 271.11 \\
      7 & 59.07 & 272.54 & 58.71 & 270.68 \\
      8 & 58.84 & 271.84 & 58.39 & 270.15 \\
      9 & 58.36 & 271.41 & 58.10 & 269.69 \\
      10 & 58.07 & 270.89 & 57.68 & 269.16 \\
      11 & 57.61 & 270.33 & 57.38 & 268.70 \\
      12 & 57.32 & 269.80 & 56.83 & 268.15 \\
      13 & 57.29 & 269.42 & 56.41 & 267.68 \\
      14 & 56.66 & 268.62 & 56.01 & 267.11 \\
      15 & 55.94 & 267.68 & 55.65 & 266.63 \\
      16 & 55.62 & 267.25 & 55.26 & 266.06 \\
      17 & 55.48 & 266.77 & 54.90 & 265.57 \\
      18 & 55.18 & 266.40 & 54.47 & 264.99 \\
      19 & 54.48 & 265.59 & 54.16 & 264.51 \\
      20 & 54.09 & 264.96 & 53.61 & 263.92 \\
      21 & 53.57 & 264.46 & 53.19 & 263.42 \\
      22 & 53.31 & 263.98 & 52.72 & 262.82 \\
      23 & 52.66 & 263.18 & 52.19 & 262.32 \\
      24 & 52.10 & 262.68 & 51.60 & 261.72 \\
      25 & 51.62 & 261.90 & 51.16 & 261.21 \\
      \bottomrule
    \end{tabular}
  \end{table}
\begin{figure*}[!t]
  \centering
  \subfloat[]{\includegraphics[width=0.33\textwidth]{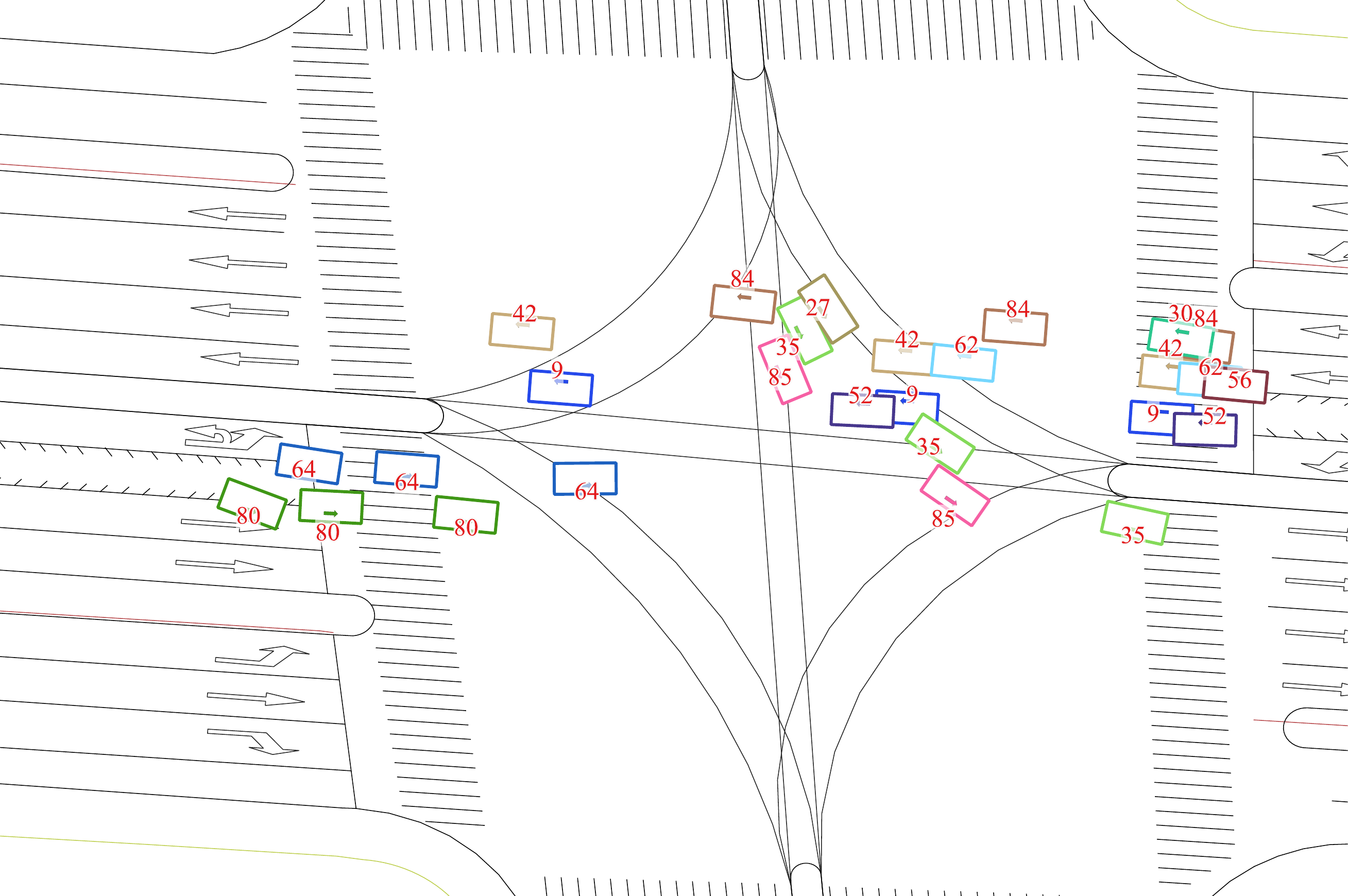}%
  \label{gl gt}}
  \hfil
  \subfloat[]{\includegraphics[width=0.33\textwidth]{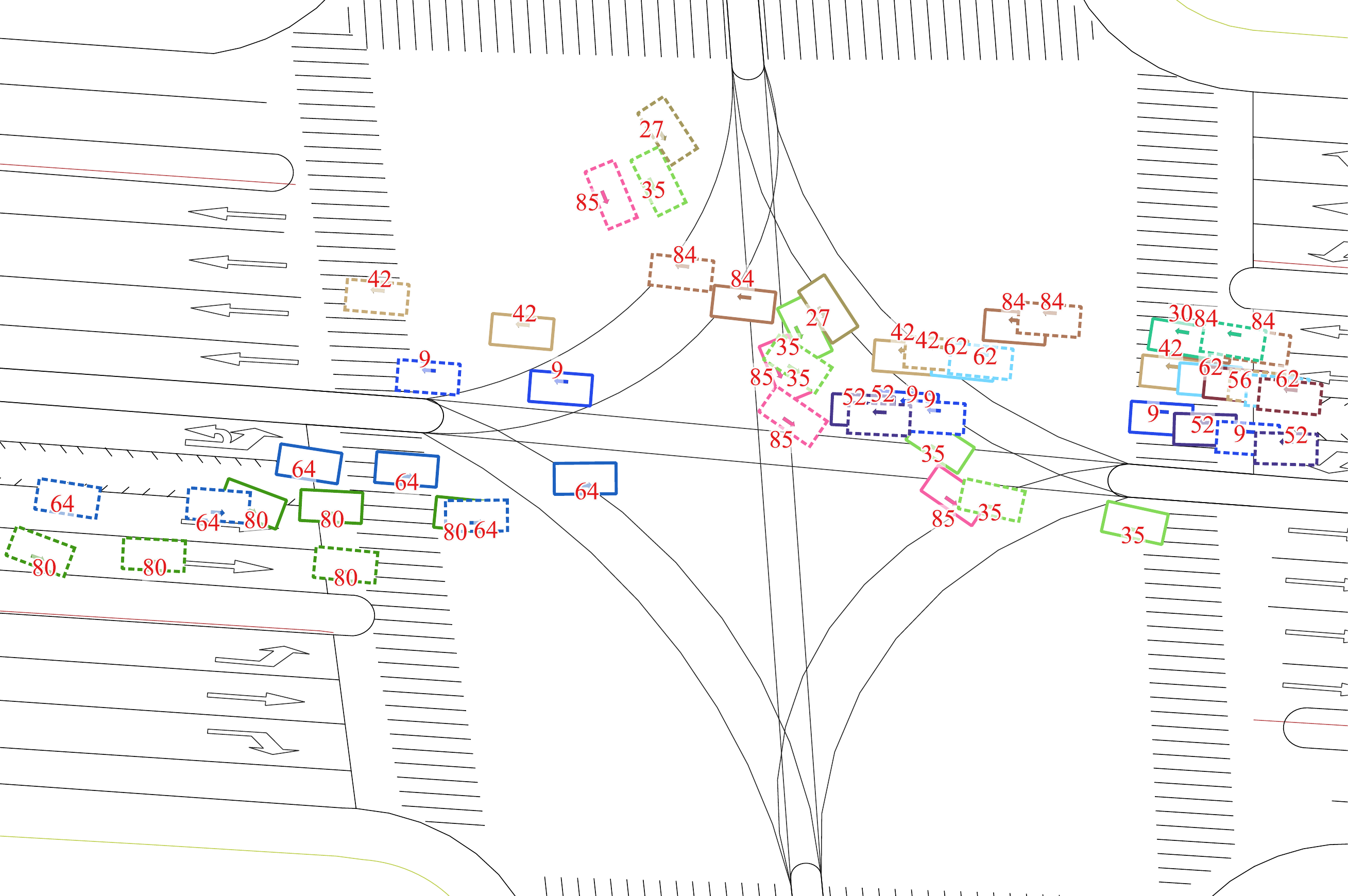}%
  \label{gl before}}
  \hfil
  \subfloat[]{\includegraphics[width=0.33\textwidth]{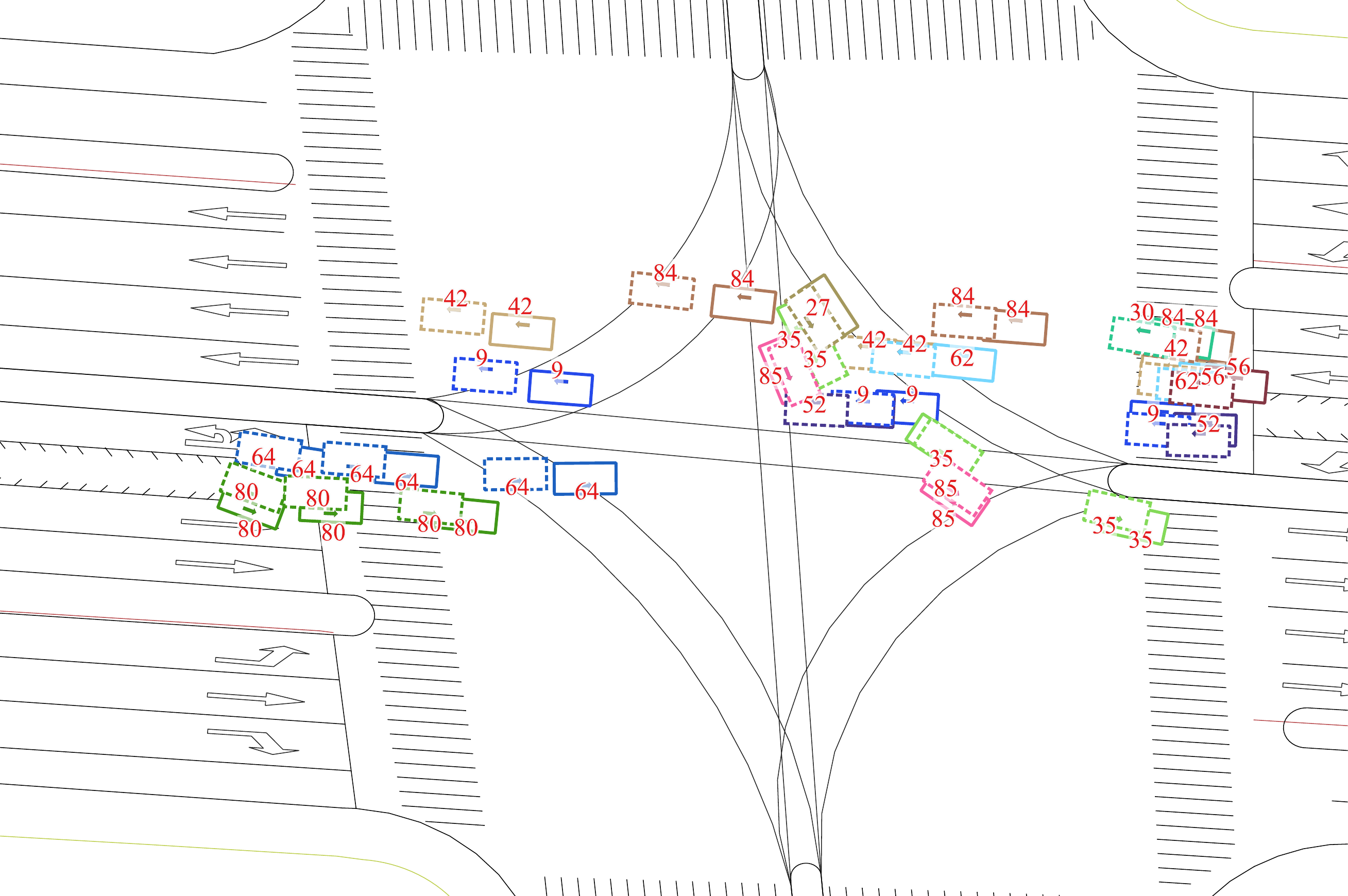}%
  \label{gl after}}
  \caption{Localization Results in Intersection 1: (a) Ground Truth, (b) Before Camera Parameter Optimization, (c) After Camera Parameter Optimization.}
  \label{jc}
  \end{figure*}
\begin{table}[!t]
  \caption{Weight Matrix and  Matching Result}
  \label{tab:ObjectMatchingResult}
  \centering
  \begin{tabular}{ccccc}
    \toprule
    \multicolumn{2}{c}{\multirow{2}{*}{\textbf{ID}}} & \multicolumn{3}{c}{\textbf{LiDAR}} \\
    \cline{3-5}
    & & 9 & 42 & 84 \\
    \midrule
    & 1 & 31.68642907 & 27.60446205 & \textbf{17.84351904} \\
    \textbf{VIDEO} & 6 & \textbf{19.32752349} & 17.20332513 & 13.44238212 \\
    & 12 & 25.31265037 & \textbf{21.23068335} & 11.46974034 \\
    \bottomrule
  \end{tabular}
\end{table}
\begin{table}[!t]
  \caption{Parameter Ranges and Results}
  \label{tab:parameter_results}
  \centering
  \begin{tabular}{cccc}
    \toprule
    \textbf{Parameter} & \textbf{Lower boundary} & \textbf{Upper boundary} & \textbf{Result} \\
    \midrule
    $f$ (pixel) & 1000 & 5000 & 3362.7385 \\
    $H$ (m) & 5 & 10 & 9.3041 \\
    $\phi$ (\textdegree) & 1 & 90 & 8.6503 \\
    $\omega_c$ (\textdegree) & 240 & 300 & 275.6085 \\
    \bottomrule
  \end{tabular}
\end{table}
\begin{table}[!b]
  \caption{Selected Parameters and RMSE Values}
  \label{tab:param_rmse}
  \centering
  \begin{tabular}{ccc}
    \toprule
    \textbf{Selected Param} & \textbf{RMSE-A (\textdegree)} & \textbf{RMSE-D (m)} \\
    \toprule
    before optimization   & 4.05             & 8.94                \\
    \hline
    $f$                     & 3.44             & 4.01                \\
    $H$                     & 4.05             & 6.97                \\
    $\varphi$                 & 4.05             & 5.75                \\
    $f + \varphi$          & 3.44             & 3.98                \\
    $f + H$               & 3.44             & 3.98                \\
    $H + \varphi$          & 4.05             & 4.15                \\
    $f + H + \varphi$      & 3.61             & 4.02                \\
    $f + H + \varphi + \omega_h$ & 3.59             & 4.06                \\
    \hline
  \end{tabular}
\end{table}
\begin{table}[!b]
  \caption{Selected Data and RMSE Values}
  \label{tab:data_rmse}
  \centering
  \begin{tabular}{ccc}
    \toprule
    \textbf{Selected Data} & \textbf{RMSE-A (\textdegree)} & \textbf{RMSE-D (m)} \\
    \toprule
    before optimization & 4.05 & 8.94 \\
    \hline
    56 & 3.64 & 7.34 \\
    62 & 3.58 & 5.19 \\
    42 & 3.43 & 4.56 \\
    56 + 62 + 42 & 3.41 & 4.18 \\
    9 + 84 + 42 & 3.42 & 4.30 \\
    35 & 5.01 & 7.79 \\
    35 + 85 + 27 & 4.74 & 8.25 \\
    Total Data & 3.59 & 4.06 \\
    \bottomrule
  \end{tabular}
\end{table}
The approximating initially involved filtering out outlier IDs generated by the fusion of LiDAR and video tracking by Algorithm \ref{algo:outlier-rejection}, as illustrated in Figure \ref{lidar video outlier}. Following the filtration, the weight matrix resulting from the Hungarian algorithm for associating video tracking IDs with tracking IDs in the LiDAR is presented in Table \ref{tab:ObjectMatchingResult}.
In Algorithm \ref{algo:outlier-rejection}, $T_{a}$, $T_{d}$, $T_{v}$, and $T_{p}$ were set at 5\textdegree, 50 m, 2 m/s, and 600 pixels, respectively.
During the optimization process, in order to converge to reasonable results, we imposed constraints on various parameters, as shown in Table \ref{tab:parameter_results}.
We conducted comparative experiments on different parameter combinations where errors were introduced to the original parameters to enhance the comparison results. 
The results from TABLE \ref{tab:param_rmse} showed that our algorithm is most affected by the camera focal length $f$, followed by pitch $\theta$, and least affected by height $H$, as indicated by the distance root mean square error (RMSE-D) and angle root mean square error (RMSE-A). 
Incorporating the approximation of the focal length reduced the angle error and improves distance accuracy for pitch $\theta$ and height $H$. 
Optimizing all three parameters together results in the highest accuracy for the localization algorithm.
One noteworthy difference in Table \ref{tab:param_rmse}, as compared to common belief, is that when optimizing in conjunction with the $\omega_h$, the outcomes may not necessarily be superior to optimizing without it. 
while the inclusion of the heading angle in the optimization slightly reduced angular errors, the reduction in distance errors is not significant, and there is even a slight increase. 
However, overall, it still contributes to the optimization effect compared to initial localization results. 
We believe that such result is to some extent related to the initial value $\phi$.

In addition to evaluating the impact of camera parameters, we investigated the influence of dataset distribution on the optimization algorithm by jointly optimizing the focal length $f$, pitch $\theta$, height $H$, and heading $\omega_h$, as presented in Fig. \ref{jc} and TABLE \ref{tab:data_rmse}. 
We observed that data distribution within specific distance ranges influenced the accuracy of the approximation. 
Testing with vehicles covering longer distance ranges led to decreased distance and angle errors. 
By selecting only straight-driving vehicles for optimization, the method effectively reduced angle errors, as it simultaneously optimizes both the distance and focal length $f$, leading to angle optimization as well. 
Additionally, even with turning vehicles, the optimization results from straight-driving vehicles sufficed to achieve accurate approximation.
To ensure a fair comparison, we analyzed the results after 30,000 epochs of optimization.
First, we selected three vehicles in the middle lane, arranged from nearest to farthest. From the optimization results of LiDAR IDs in the Table \ref{tab:data_rmse} and the Fig. \ref{jc}, it's evident that optimizing the trajectories of more distant vehicles leads to more noticeable error reduction. 
The combination of all three cars yields the most accurate results.
Next, we compared the results of three vehicles on three lanes to those of three vehicles on a single lane. 
Surprisingly, the former didn't outperform the latter.
We attribute this, to a large extent, to interference from noise data points. 
In scenarios where multiple vehicles are present, the LiDAR-based localization results are inevitably affected by factors such as occlusions, leading to some degree of localization error.
Furthermore, we conducted experiments with individual turning vehicles and three turning vehicles, revealing that while combining three turning vehicles results in lower angular errors than a single turning vehicle, the distance error actually increases, and optimizing the trajectories of three turning vehicles doesn't produce better results than optimizing a single straight-line vehicle. 
We attribute this primarily to the data distribution.

\begin{figure*}[!t]
  \centering
  \subfloat[]{\includegraphics[width=0.32\textwidth]{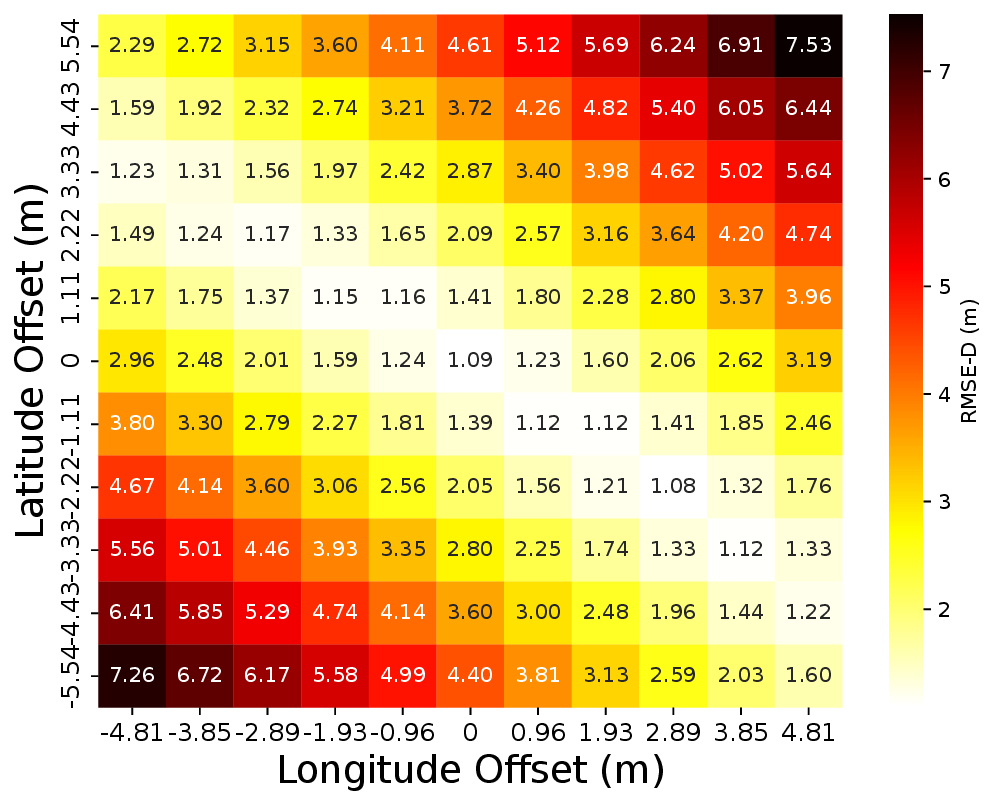}%
  \label{distance error}}
  \hfill
  \subfloat[]{\includegraphics[width=0.33\textwidth]{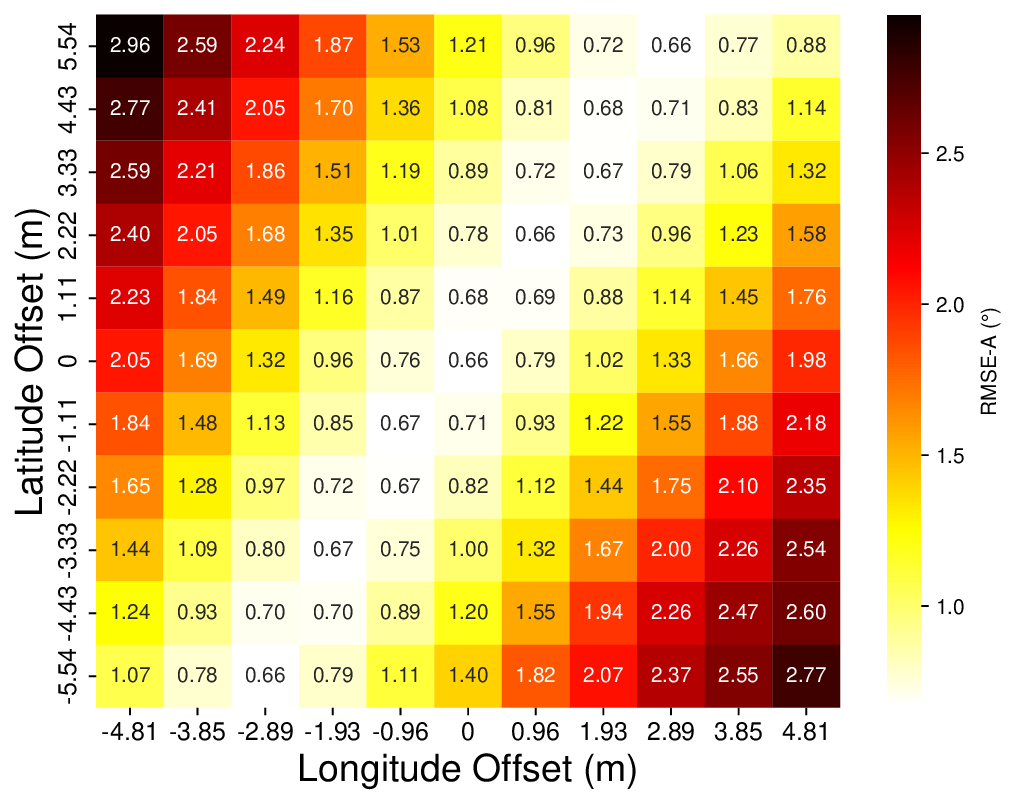}%
  \label{angle error}}
  \hfill
  \subfloat[]{\includegraphics[width=0.32\textwidth]{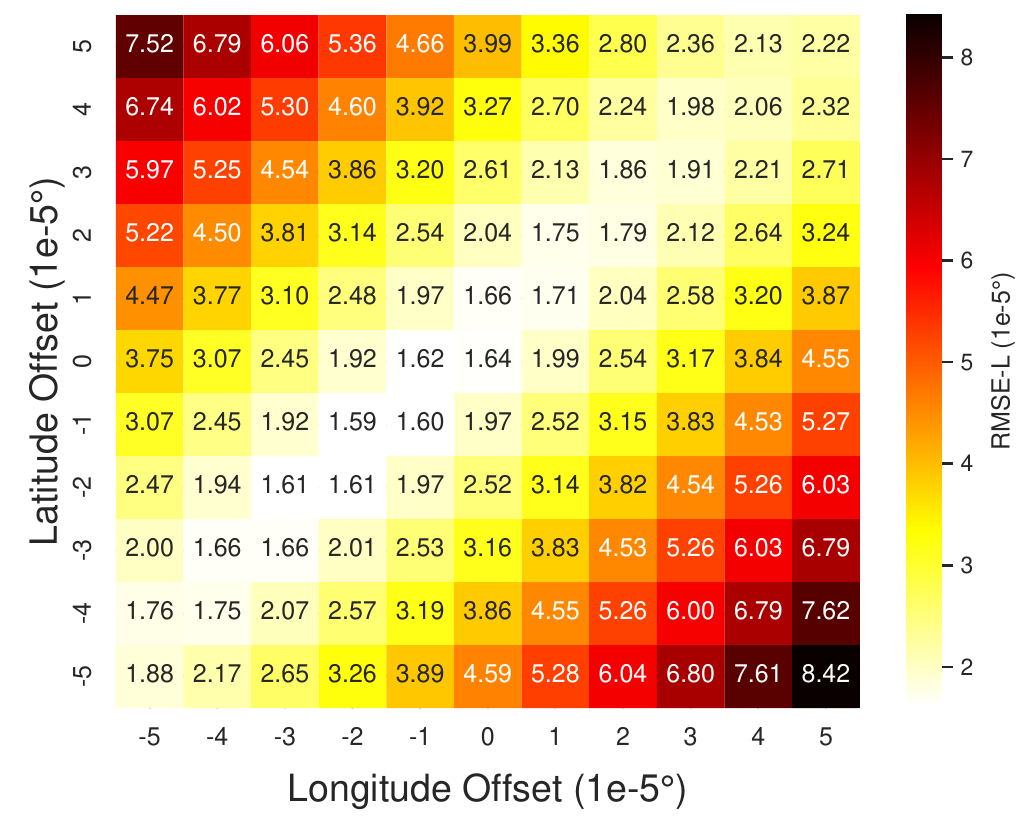}%
  \label{location error}}
  \hfill
  \caption{Impact of Latitude and Longitude Errors of Camera on Distance and Angle Estimation: (a) RMSE-D Result, (b) RMSE-A Result, (c) RMSE-L Result.}
  \label{camera lo_la_error}
\end{figure*}
\begin{figure*}[!t]
  \begin{minipage}{0.5\textwidth}
  \centering
  \subfloat[]{\includegraphics[width=0.5\columnwidth]{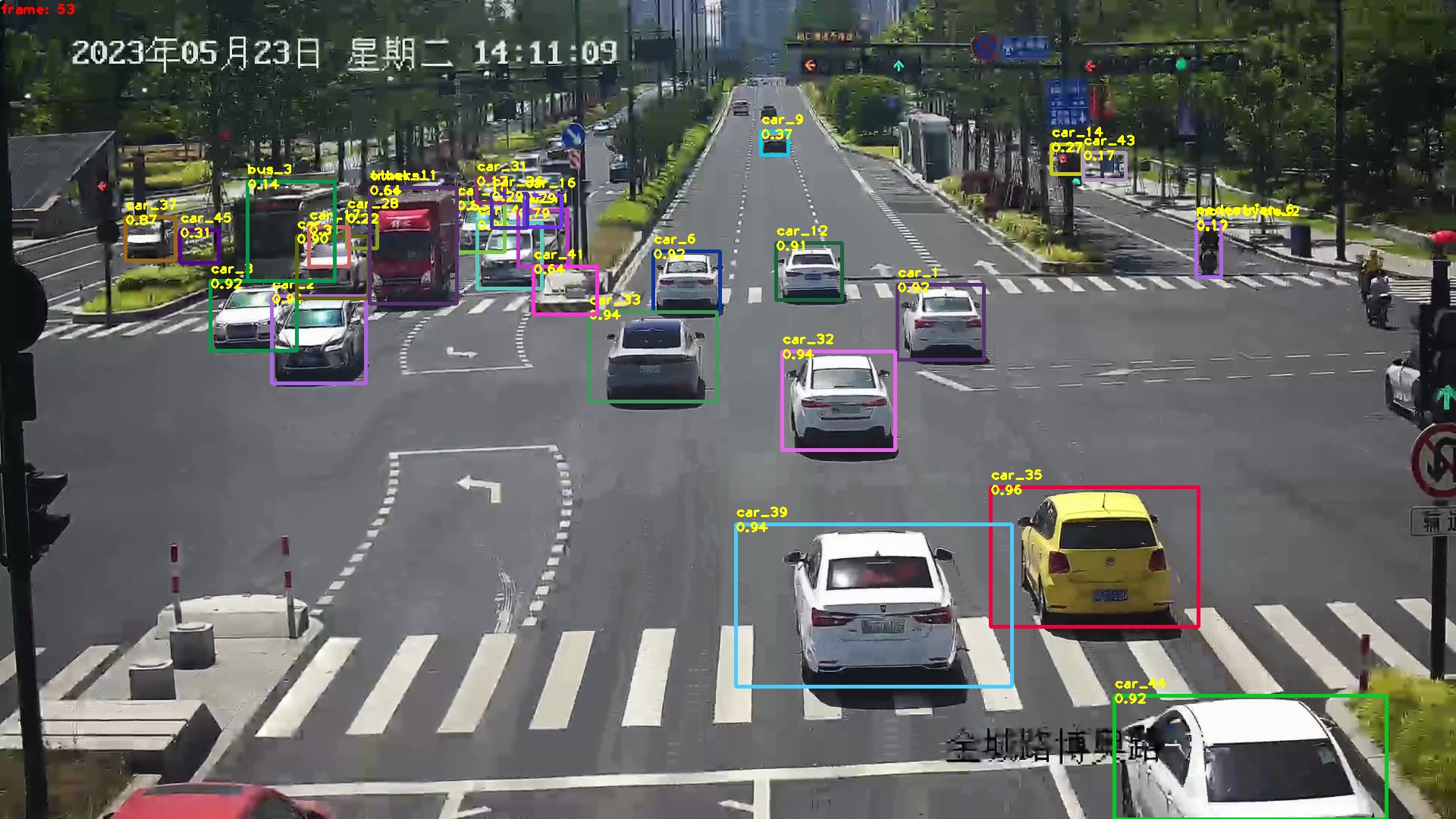}%
  \label{E}}
  \hfill
  \subfloat[]{\includegraphics[width=0.5\columnwidth]{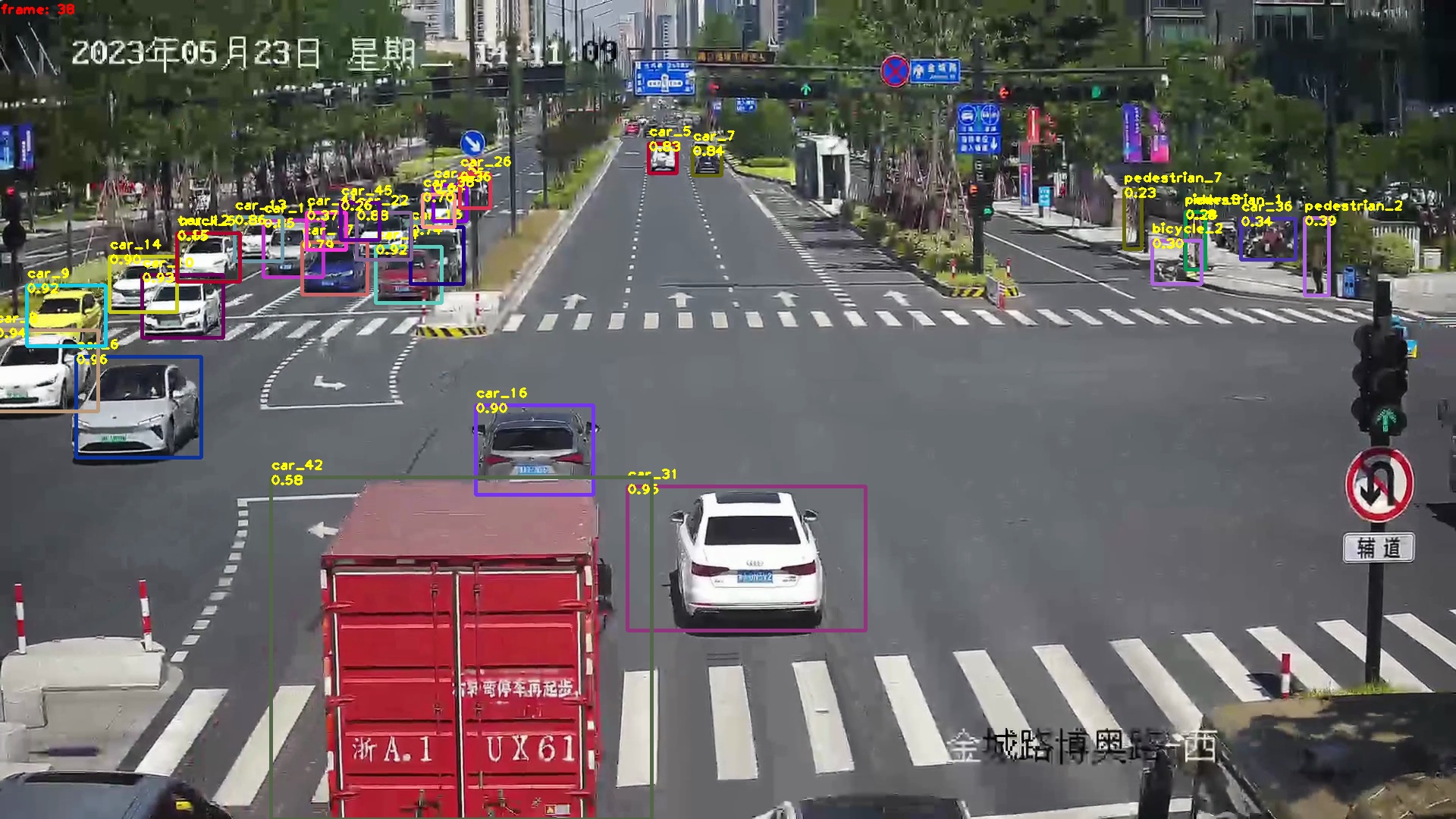}%
  \label{W}}
  \hfill
  \subfloat[]{\includegraphics[width=0.5\columnwidth]{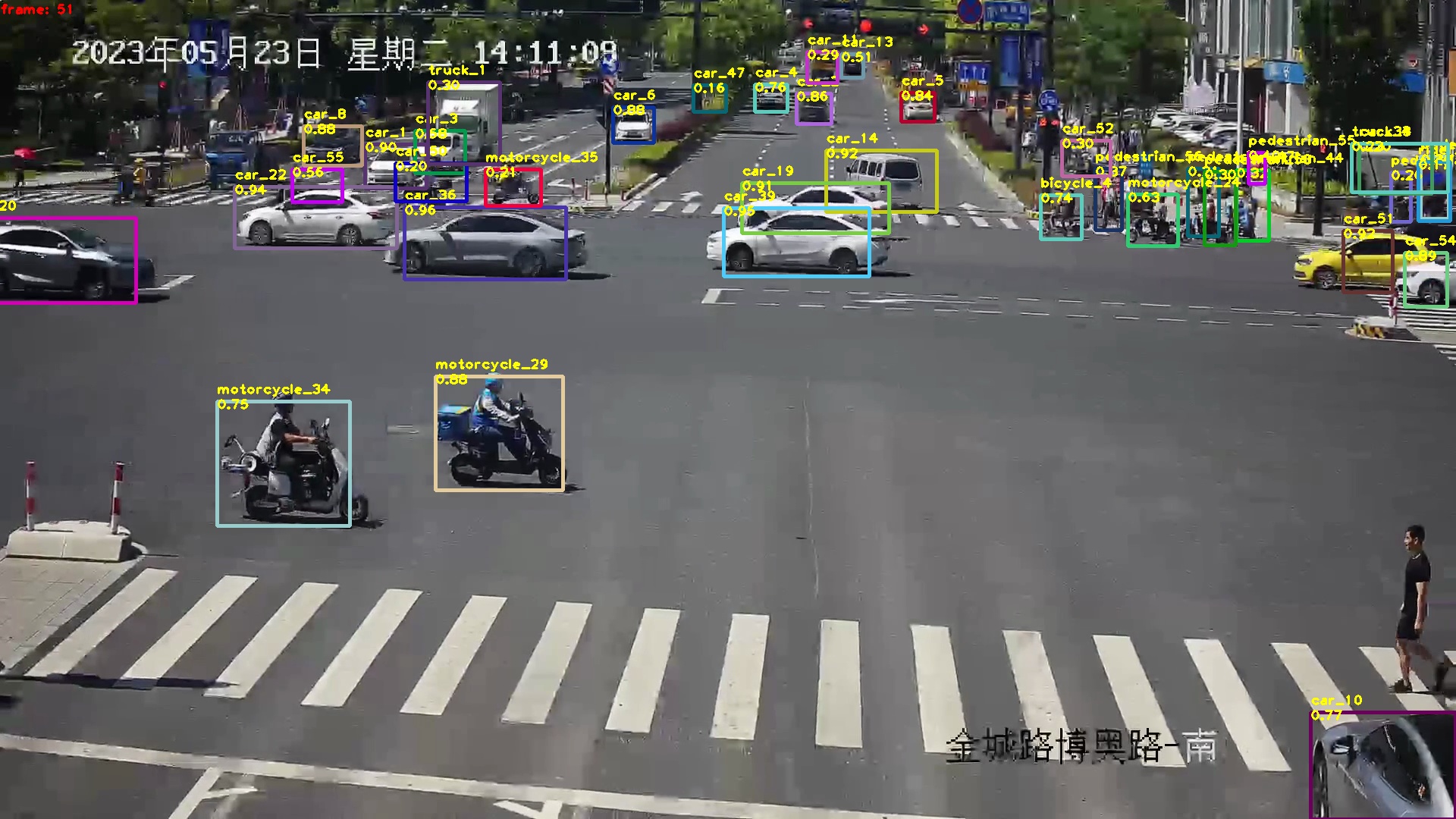}%
  \label{S}}
  \hfill
  \subfloat[]{\includegraphics[width=0.5\columnwidth]{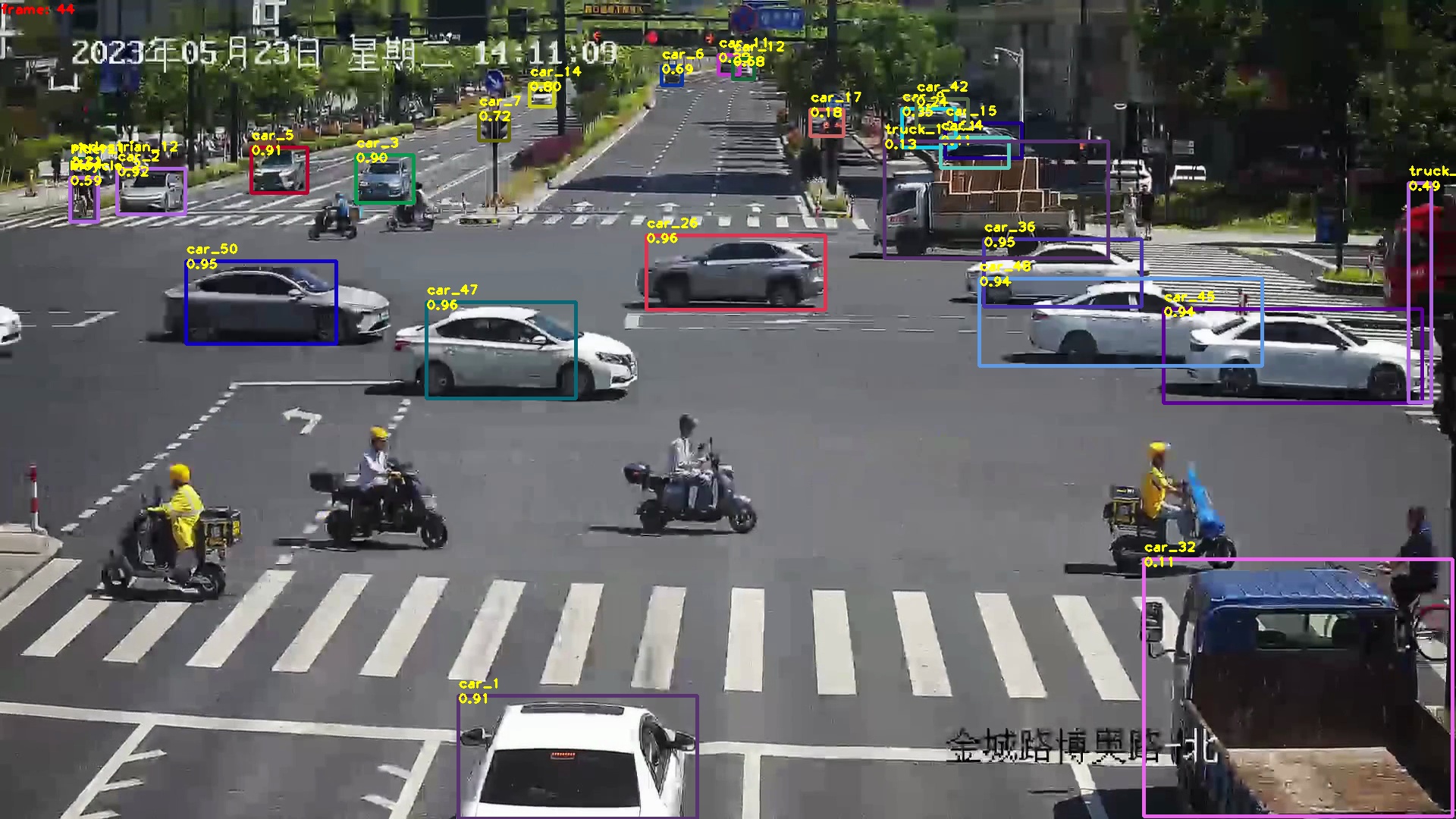}%
  \label{N}}
  \hfill
  \label{fusion approximation_ns}
  \end{minipage}%
  \begin{minipage}{0.5\textwidth}
    \centering
    \includegraphics[width=\columnwidth]{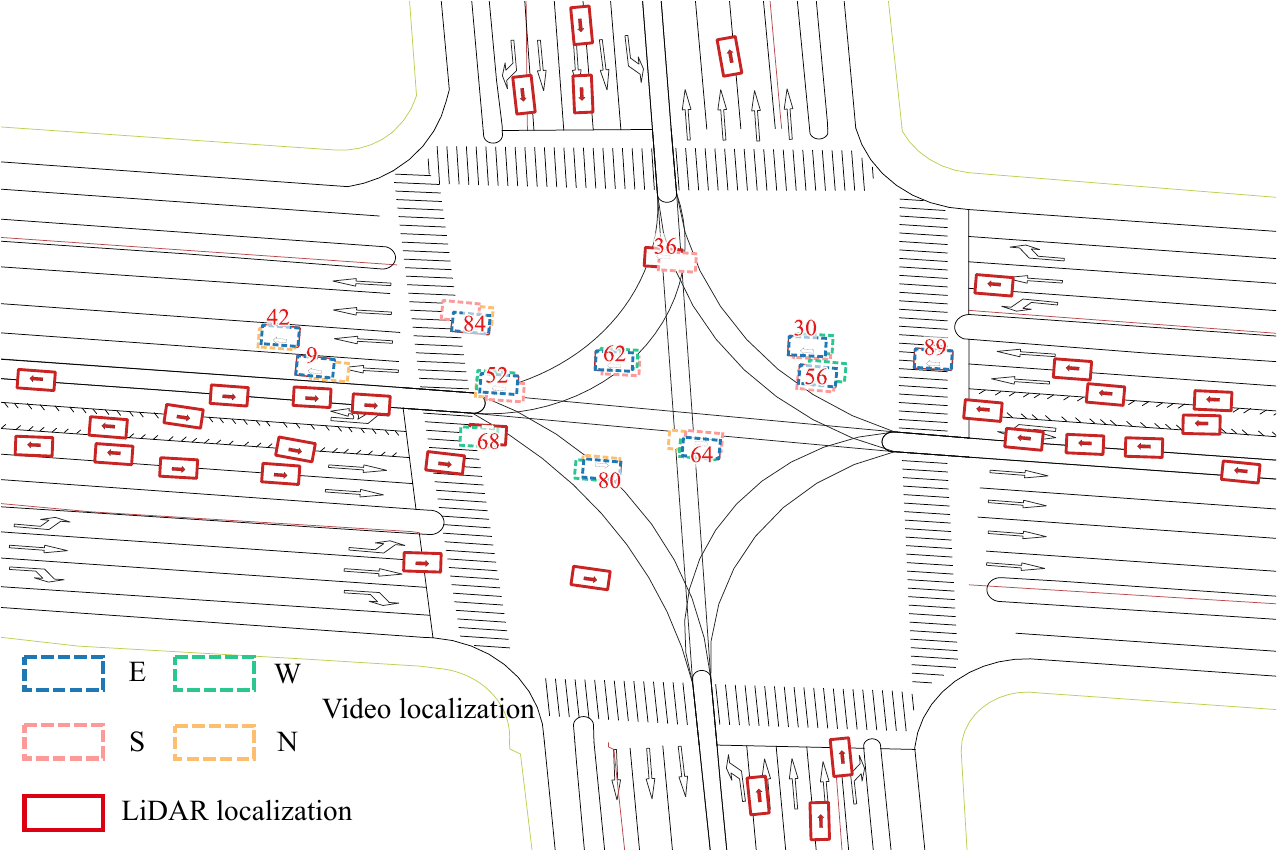}
    \label{syn_all}
  \end{minipage}%
  \caption{Multi-view, multi-sensor spatial synchronization results. In the left panel, (a), (b), (c), and (d) represent the results of four vehicles tracked by video, corresponding to the four cardinal directions: east, west, south, and north. The right panel shows the spatial synchronization results in BEV (Bird's Eye View).}
  \label{syn}
\end{figure*}

Considering the camera's latitude and longitude errors during deployment, we simulated different error scenarios and observed an increase in distance and angle estimation errors as errors grew as shown in Fig. \ref{camera lo_la_error}\ref{sub@distance error} and Fig. \ref{camera lo_la_error}\ref{sub@angle error}. The trends differed significantly, with angle errors deviating less in the heading angle direction, and distance errors deviating less in the direction perpendicular to the heading angle. Overall, the distance error is larger than the angle error. 
However, in order to analyze the impact of angle and distance errors on the final localization error more clearly, we conducted an analysis of the localization error in Fig. \ref{camera lo_la_error}\ref{sub@location error}. It can be observed that the localization error is closer to the angle error.
So, For spatial synchronization purposes, a 2 m camera mounting position error range is acceptable, while for pure localization, a wider acceptable error range can be considered.
\subsection{Synchronization Evaluation}
\begin{table}[!t]
  \caption{The Comparison of the Two Implementations in Terms of Operation Time and Accuracy}
  \label{tab:comparison_of_the_two_implementations}
  \centering
  \begin{tabular}{ccc}
    \toprule
    \textbf{Methods} & \textbf{Operation Time (Minutes)} & \textbf{Relative Accuracy (\%)} \\
    \toprule
    CST & 28 & 53.5 \\ 
    PST & 3 & 85.6 \\
    \bottomrule
  \end{tabular}
\end{table}
To evaluate the applicability of our framework in a multi-camera scenario, we selected an intersection scene monitored by two cameras, as shown in the Fig. \ref{syn}. The majority of vehicles were detected simultaneously by four cameras, while a few vehicles were not detected and localized by the other camera due to the different camera's fov. The localization results of the four cameras were compared with the LiDAR localization on the HD map. 
It can be observed that the localization results of four cameras closely match the results from the LiDAR localization, demonstrating the effectiveness of our framework in multi-camera scenarios.

In order to make a clear comparison with the previous methods, we manually performed camera calibration and spatial synchronization, following the procedures described in \cite{Zheng2022ARS}. Experimental comparisons were conducted as shown in Table \ref{tab:comparison_of_the_two_implementations}, and it is evident that our approach exhibits significant advantages in terms of deployment efficiency and accuracy.
\section{DISSCUSSION}\label{discussion}
This paper provides a perspective for the subsequent research where GT (Ground Truth) calibration parameters for the camera may not be essential, especially for monocular localization and spatial synchronization.
Obtaining the GT of camera parameters is challenging, when combined with camera global parameters (e.g., camera mounting height, initial latitude and longitude, and camera heading angle) may not necessarily yield the highest precision in monocular localization and spatial synchronization. 
This is due to the possibility of errors in the global camera parameters, and variations in localization results obtained through different localization formulas.
As a result, there may not be a genuine need for actual camera parameters or an extremely accurate camera calibration algorithm. 
Optimizing the results of camera calibration alongside global parameters may be a favorable approach.

In the case of Algorithm \ref{algo:vehicle-selection-heading}, adjustments may be necessary when applied to expressway or scenarios with time asynchrony. 
The estimation of the heading angle relies on the assumption of camera's orientation are roughly aligned with the direction of the road and the provision of vehicle's start-up time. 
Vehicle starting times are readily obtainable in urban traffic scenarios, while the assumption that the camera's orientation aligns with the road's direction may not be easily met in some expressway scenarios.

While various spatial synchronization methods have been proposed, we believe that one of the most pressing and crucial issues currently is the lack of a clear evaluation standard to assess the accuracy of spatial synchronization. 
A comprehensive evaluation standard would enable precise experimental comparisons between different methods and provide a reference for more reliable deployments. 
We intend to work towards addressing this gap in the future.
And, although the framework presented in this paper is laser-based, it can be further combined with V2I (Vehicle-to-Infrastructure) and ego GNSS data to reduce costs, similar to the approach in \cite{Ojala2022InfrastructureCC}, for the deployment of more cost-effective spatial synchronization models.
While our method significantly reduces manual intervention, addressing the challenge of matching due to differences in camera and LiDAR installation positions and detection ranges, our algorithm incorporates several threshold settings. This means that these thresholds may need adjustment in different scenarios. In the future, we plan to introduce simpler algorithms to optimize this aspect and further simplify the deployment process.
Although the optimization algorithm can mitigate the effects of spatial asynchrony to some extent, significant time synchronization errors can lead to lower convergence accuracy. We also aim to explore more comprehensive spatio-temporal synchronization algorithms and design additional experiments to investigate the impact of time synchronization errors on optimization algorithms.
Finally, the system proposed in this paper can also be easily integrated with MTMC (Multi-Target Multi-Camera) tracking, combined with ReID (Re-identification) algorithms, to achieve even faster spatial synchronization.
This is also a interesting research direction in the future.
\section{CONCLUSION}\label{conclusion}
In this paper, we present a spatial synchronization framework based on PST, aligning the coordinate systems of multiple sensors in parallel with a global coordinate system. 
Our framework is tested in real-world scenarios, and the experimental results indicated that it achieved high precision in spatial synchronization, tolerating certain a certain margin of errors in camera installation latitude, longitude, and height. 
Meanwhile, compared to approach based CST, our framework significantly reduces manual intervention and the associated accumulation of errors, achieves nearly tenfold speed in automatically realizing spatial synchronization from multiple viewpoints. 
Our framework is tailored for large-scale, multi-view, multi-sensor spatial synchronization scenarios, taking into account practical challenges such as potential changes in camera parameters, different sensor installation placements, orientations, height and diverse deployment algorithms in real-world scenarios, providing low maintenance costs and easy scalability.
\bibliographystyle{IEEEtran}
\bibliography{reference}

\end{document}